\documentclass[preprint,aps,12pt,notitlepage,nofootinbib,tightenlines]{revtex4}
\usepackage{setspace}
\usepackage{placeins}
\usepackage{amsmath}
\usepackage{caption}
\usepackage{subcaption}
\usepackage[per-mode=symbol]{siunitx}
\usepackage{booktabs}
\usepackage{multirow}
\usepackage{graphicx}
\usepackage{enumitem}
\usepackage{underscore}
\usepackage{bm}
\usepackage{times}
\usepackage{braket}
\usepackage{color}
\usepackage{epsfig}
\usepackage{slashed}
\usepackage{hyperref}
\usepackage{array}
\usepackage{float}
\newcommand{\beq}{\begin{eqnarray}}
\newcommand{\eeq}{\end{eqnarray}}
\newcommand{\be}{\begin{equation}\begin{aligned}}
\newcommand{\ee}{\end{aligned}\end{equation}}

\definecolor{Red}{rgb}{1.,0.,0.}

\definecolor{Blue}{rgb}{0.,0.,1.}

\definecolor{nicered}{rgb}{0.7,0.1,0.1}
\definecolor{nicegreen}{rgb}{0.1,0.5,0.1}
\def\lsim{ {\ \lower-1.2pt\vbox{\hbox{\rlap{$<$}\lower6pt\vbox{\hbox{$\sim$}}}}\ } }
\def\gsim{ {\ \lower-1.2pt\vbox{\hbox{\rlap{$>$}\lower6pt\vbox{\hbox{$\sim$}}}}\ } }
\hypersetup{colorlinks,citecolor=nicegreen,linkcolor=nicered}
\begin{document}
\title{Probing the Single Production of First-Generation Singlet Vector-like Leptons at Future $e^+e^-$ Colliders}
\author{Yao-Bei Liu$^{1,2}$\footnote{E-mail: liuyaobei@hist.edu.cn}, Stefano Moretti$^{3,4}$\footnote{E-mail: stefano.moretti@cern.ch} }
\affiliation{1. Henan Institute of Science and Technology, Xinxiang 453003, China\\
2. School of Electro-Mechanical Engineering, Zhongyuan University of Science and Technology, Xuchang 461000, China\\
3. School of Physics and Astronomy, University of Southampton, Highfield, Southampton SO17 1BJ, United Kingdom \\
4. Department of Physics and Astronomy, Uppsala University, Box 516, 751 20 Uppsala, Sweden}
\vspace*{0.5truecm}
\begin{abstract}
{\small \begin{singlespace}
We study the single production of first-generation weak-isosinglet vector-like leptons (VLLs) at future $e^+e^-$ colliders through $e^+e^-\to e^\pm E^\mp$, with the subsequent decays $E^\pm\to W^\pm\nu_e$, $e^\pm Z$, and $e^\pm h$. For the heavy VLL masses considered, the hadronic decays of the $W$ and $Z$ bosons are reconstructed as fat-jets, while the Higgs boson is reconstructed from two resolved $b$-tagged jets at $\sqrt{s}=1~{\rm TeV}$ and as a fat-jet at $\sqrt{s}=1.5~{\rm TeV}$. Using detector-level Monte Carlo simulations for the 1~TeV ILC and 1.5~TeV CLIC, we perform a baseline cut-based analysis in which the well-defined two-body production kinematics are exploited through a mass-hypothesis-dependent selection on the associated-electron energy. At fixed nominal center-of-mass energy, this observable is in one-to-one correspondence with the VLL recoil mass and therefore provides a common mass-sensitive discriminator for all three decay channels. We derive the $2\sigma$ exclusion and $5\sigma$ discovery reaches on the VLL--electron mixing angle $\sin\theta_L$ as functions of the integrated luminosity. For the benchmark points studied, sizeable sensitivity is retained at $m_E=900~{\rm GeV}$ at the ILC and $m_E=1400~{\rm GeV}$ at CLIC, while the accessible values of $\sin\theta_L$ extend below the reference electroweak precision bound used in this study. These results demonstrate the complementary potential of direct searches for first-generation singlet VLLs at future lepton colliders.
\end{singlespace}}
\end{abstract}

\maketitle
\newpage
\section{Introduction}

Vector-like leptons (VLLs) constitute a well-motivated class of
beyond-the-Standard-Model (BSM) scenarios and arise in a variety of
theoretical frameworks, including supersymmetry~\cite{Graham:2009gy,Endo:2011xq,Martin:2012dg,Endo:2012cc,Fischler:2013tva},
extra-dimensional models~\cite{Huang:2012kz,Kong:2010qd},
composite Higgs models~\cite{DeCurtis:2018iqd}, and other extensions
of the Standard Model (SM)~\cite{He:1999vp,Wang:2013jwa,He:2001fz,He:2014ora}.
They are also phenomenologically interesting because they can play a
role in addressing several open problems in particle physics, including
the hierarchy problem through new dynamics in the Higgs
sector~\cite{Arkani-Hamed:2012dcq}, the stabilization of the
electroweak vacuum~\cite{Xiao:2014kba,Cingiloglu:2024vdh}, and the
realization of viable dark matter candidates~\cite{Schwaller:2013hqa,Halverson:2014nwa,Bahrami:2016has,Bhattacharya:2018fus}.
In addition, VLLs can give sizeable contributions to the anomalous
magnetic moments of the electron and muon~\cite{Hiller:2019mou,DeJesus:2020yqx,Frank:2020smf,Dermisek:2021ajd,Brune:2022rlo,Guedes:2022cfy,Erdelyi:2025axy,Athron:2025ets},
as well as to the $W$-boson mass~\cite{He:2022zjz,Kawamura:2022fhm},
making them relevant to the interpretation of these precision
measurements.

A defining feature of VLLs is their vector-like nature, namely that
their left- and right-handed components transform identically under the
SM gauge group. This allows gauge-invariant bare mass terms independent
of electroweak symmetry breaking, in contrast to the masses of chiral
SM fermions generated through the Higgs mechanism. Consequently, VLLs
can be subject to comparatively weaker constraints from electroweak
precision observables~\cite{delAguila:2008pw,Ishiwata:2013gma} and
Higgs coupling measurements~\cite{Kearney:2012zi,Altmannshofer:2013zba,Falkowski:2013jya}.
At the same time, they give rise to a rich variety of collider
signatures that have been extensively studied in the
literature~\cite{Ellis:2014dza,Ishiwata:2015cga,Dermisek:2014cia,Crivellin:2020ebi,Endo:2020tkb,Chakrabarty:2020jro,Guedes:2021oqx,Raju:2022zlv,Li:2023mrw,Dermisek:2014qca,Dermisek:2015oja,Kumar:2015tna,Dermisek:2015hue,Chen:2016lsr,Kawamura:2019rth,Freitas:2020ttd,OsmanAcar:2021plv,Kawamura:2021ygg,Bonilla:2021ize,Baspehlivan:2022qet,Cao:2023smj,Bernreuther:2023uxh,Kawamura:2023zuo,Bigaran:2023ris}.

Depending on the SM lepton generation with which they mix, VLLs can
be associated with the electron, muon, or tau sectors. In this work,
we focus on the first-generation scenario and denote the charged VLL
by $E^{\pm}$. This case exhibits phenomenological features distinct
from those of its second- and third-generation counterparts. At the
Large Hadron Collider (LHC), direct searches have predominantly
targeted pair-produced VLLs decaying into leptons and electroweak gauge
bosons~\cite{ATLAS:2015qoy,ATLAS:2023sbu,CMS:2019hsm,CMS:2022nty,CMS:2022cpe,CMS:2024bni}.
The most recent ATLAS analysis~\cite{ATLAS:2024mrr} sets 95\%
confidence-level (CL) lower limits on the masses of vector-like
electrons (muons), excluding doublet representations up to
1220~GeV (1270~GeV) and singlet representations up to
320~GeV (400~GeV). These results illustrate the particular challenge
of probing singlet VLLs in the hadronic environment of the LHC.

The sizable quantum chromodynamics (QCD) backgrounds at hadron
colliders pose an additional challenge for weak-isosinglet VLL
searches~\cite{Bhattiprolu:2019vdu}, motivating complementary studies
at cleaner $e^+e^-$ machines such as the International Linear Collider
(ILC)~\cite{ILC:2013jhg,ILCInternationalDevelopmentTeam:2022izu} and
the Compact Linear Collider (CLIC)~\cite{CLICDetector:2013tfe,Franceschini:2019zsg}.
The clean experimental environment and well-defined initial-state
kinematics of lepton colliders provide important advantages for searches
for new weakly interacting states. VLL production at lepton colliders
has consequently been investigated in both pair-production
channels~\cite{Yang:2021dtc,Shang:2021mgn,Bhattacharya:2021ltd,Shang:2023rfv,Bhattiprolu:2023yxa,Yue:2024sds,Shen:2025mxe,Sagheer:2026eql,Liu:2025ori,Liu:2025akp}
and single-production channels~\cite{Das:2020gnt,Das:2020uer,Yue:2024ftz,Yue:2025xvk,Li:2025epjc}.
While pair production is mainly limited by the kinematic threshold and
is largely independent of the mixing parameter $\epsilon$, the
single-production cross section scales approximately as $\epsilon^2$.
Single production therefore provides direct sensitivity to the mixing
between the VLL and the corresponding SM lepton. In the singlet model,
$\epsilon$ is related to the left-handed mixing angle through
$\sin\theta_L \simeq \epsilon v/m_E$. For comparison with electroweak precision constraints, we adopt
$\sin\theta_L\lesssim0.021$ at 95\% CL from Ref.~\cite{deBlas:2013gla} as a
reference bound. Recent studies of single production of singlet VLLs
mixing with third-generation leptons at the LHC find sensitivity only
to masses below about $500~{\rm GeV}$, even when advanced
machine-learning techniques are employed~\cite{Cui:2026wwo}.

In this work, we investigate the single production of first-generation
singlet VLLs at the ILC with $\sqrt{s}=1$~TeV and at CLIC with
$\sqrt{s}=1.5$~TeV. We consider the three dominant decay channels
\[
\begin{split}
{\rm (i)}\quad&
e^+e^-\to e^{\pm}E^{\mp},
\qquad
E^{\pm}\to W^{\pm}\nu_e,
\qquad
W^{\pm}\to q\bar q',
\\
{\rm (ii)}\quad&
e^+e^-\to e^{\pm}E^{\mp},
\qquad
E^{\pm}\to e^{\pm}Z,
\qquad
Z\to q\bar q,
\\
{\rm (iii)}\quad&
e^+e^-\to e^{\pm}E^{\mp},
\qquad
E^{\pm}\to e^{\pm}h,
\qquad
h\to b\bar b.
\end{split}
\]

For the VLL mass range considered, $m_E=800$--$1400$~GeV, the
hadronic decays of the boosted $W$ and $Z$ bosons are reconstructed
as single fat-jets. In the Higgs channel, the Higgs decay products are
reconstructed as two resolved $b$-tagged jets at
$\sqrt{s}=1$~TeV, while a fat-jet reconstruction is employed at
$\sqrt{s}=1.5$~TeV. A common mass-hypothesis-dependent selection is
constructed from the energy of the electron produced in association
with the VLL. For a fixed nominal center-of-mass energy, the
associated-electron energy is in one-to-one correspondence with the
recoil mass of the heavy VLL, so that the corresponding energy window
is kinematically equivalent to a recoil-mass window. This provides a
common way to exploit the well-defined two-body production kinematics
in all three decay channels.

We perform detailed signal and background simulations and derive the
$2\sigma$ exclusion and $5\sigma$ discovery reaches on
$\sin\theta_L$ in the $(\mathcal{L},\sin\theta_L)$ plane for
representative VLL mass benchmark points. The resulting sensitivities are compared with the reference
electroweak precision constraint adopted from Ref.~\cite{deBlas:2013gla} and with
existing collider limits, allowing us to assess the complementary
reach of direct VLL searches at future $e^+e^-$ colliders.

The remainder of this paper is organized as follows.
In Sec.~\ref{sec:model}, we introduce the singlet VLL model and discuss
the relevant interactions and single-production cross sections.
Section~\ref{sec:analysis} presents the collider simulation framework,
event reconstruction and selection strategy, including the recoil-mass
relation and the corresponding associated-electron energy selection,
as well as the resulting sensitivity projections.
Finally, Sec.~\ref{sec:conclusion} summarizes our main results and
discusses their implications for future $e^+e^-$ colliders.

\section{Theory}
\label{sec:model}

\subsection{Model Setup}

Our analysis is based on a minimal extension of the SM through the introduction of a weak-isosinglet VLL, denoted as $E^{\pm}$, which mixes with the first-generation SM leptons. Under the $SU(3)_C \times SU(2)_L \times U(1)_Y$ gauge group, the new fermion transforms as
\begin{equation}
E_{L},E_{R}\sim (\mathbf{1},\mathbf{1},-1),
\end{equation}
with both chiral components carrying identical gauge charges, thereby permitting a gauge-invariant Dirac mass term.

The mass terms and mixing structure involving the VLL and the SM electron can be described by the following Lagrangian in two-component spinor notation:
\begin{equation}
-\mathcal{L} = m_E E\tilde{E} + \epsilon H L_e \tilde{E} + y_e H L_e \tilde{e} + \text{c.c.},
\end{equation}
where $H$ denotes the SM Higgs doublet, $L_e = (e, \nu_e)_L$ is the first-generation lepton doublet in the gauge basis, $y_e$ is the SM electron Yukawa coupling, and $\epsilon$ is the Yukawa coupling that controls the mixing. The two-component fields $E$ and $\tilde{E}$ correspond to the left- and right-handed components of the vector-like electron, respectively.

Upon electroweak symmetry breaking, the charged-lepton mass matrix in the $(e, E)$ basis takes the form
\begin{equation}
\mathcal{M} =
\begin{pmatrix}
y_e v & \epsilon v\\
0 & M
\end{pmatrix}.
\end{equation}
with $v \simeq 174$ GeV being the Higgs vacuum expectation value. The corresponding mass eigenvalues are given by
\begin{equation}
m_e \simeq y_e v \left(1 - \frac{\epsilon^2 v^2}{2M^2} + \cdots\right), \qquad
m_E \simeq M \left(1 + \frac{\epsilon^2 v^2}{2M^2} + \cdots\right).
\end{equation}
The mixing between the SM electron and the VLL is quantified by a small mixing angle $\theta_L$, which at leading order satisfies
\begin{equation}
\label{eq:mixing}
\sin\theta_L \simeq \frac{\epsilon v}{m_E}.
\end{equation}

We now discuss the phenomenological constraints on the mixing angle $\theta_L$, which is intimately related to $\epsilon$ through Eq.~\eqref{eq:mixing}. Electroweak precision data place stringent bounds on this parameter
~\cite{deBlas:2013gla,Cynolter:2008ea,Adhikary:2024esf,deBlas:2025pco,Daberstiel:2026wxe}. In the numerical comparisons presented below, we adopt
the 95\% CL bound
$\sin\theta_L\lesssim0.021$
from Ref.~\cite{deBlas:2013gla} as a reference constraint for the first-generation
mixing scenario.
For values of $\theta_L$ significantly below this limit, the VLL lifetime can become sufficiently long to give rise to displaced signatures, which are subject to dedicated searches for long-lived charged particles~\cite{Cao:2023smj,Bernreuther:2023uxh}. In the present work, we restrict our attention to the prompt-decay regime, where $\theta_L$ is chosen to be consistent with precision constraints while still being large enough to ensure that the VLL decays promptly within the detector.

\subsection{Single VLL Production at $e^+e^-$ Colliders}
\begin{figure}[h]
\centering
\includegraphics[width = 14cm ]{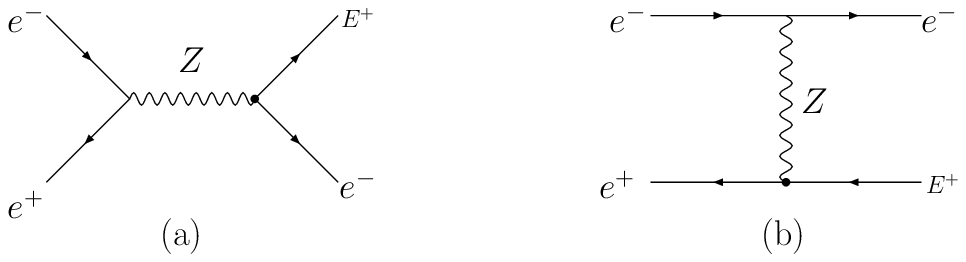}
\vspace{-14cm}
\caption{Representative Feynman diagrams for the single production of singlet first-generation VLLs via the process $e^+e^- \to e^-E^+$. }
\label{fig1}
\end{figure}

The single production of singlet VLLs proceeds via both \(s\)-channel and \(t\)-channel \(Z\)-boson exchanges, with representative Feynman diagrams shown in Fig.~\ref{fig1}. The corresponding charge-conjugate processes are included in the numerical calculation. For fixed collider energy and VLL mass, the production cross section is proportional to $\epsilon^2$.

\begin{figure}[h]
\begin{center}
\centerline{\hspace{1.0cm}\epsfxsize=9cm\epsffile{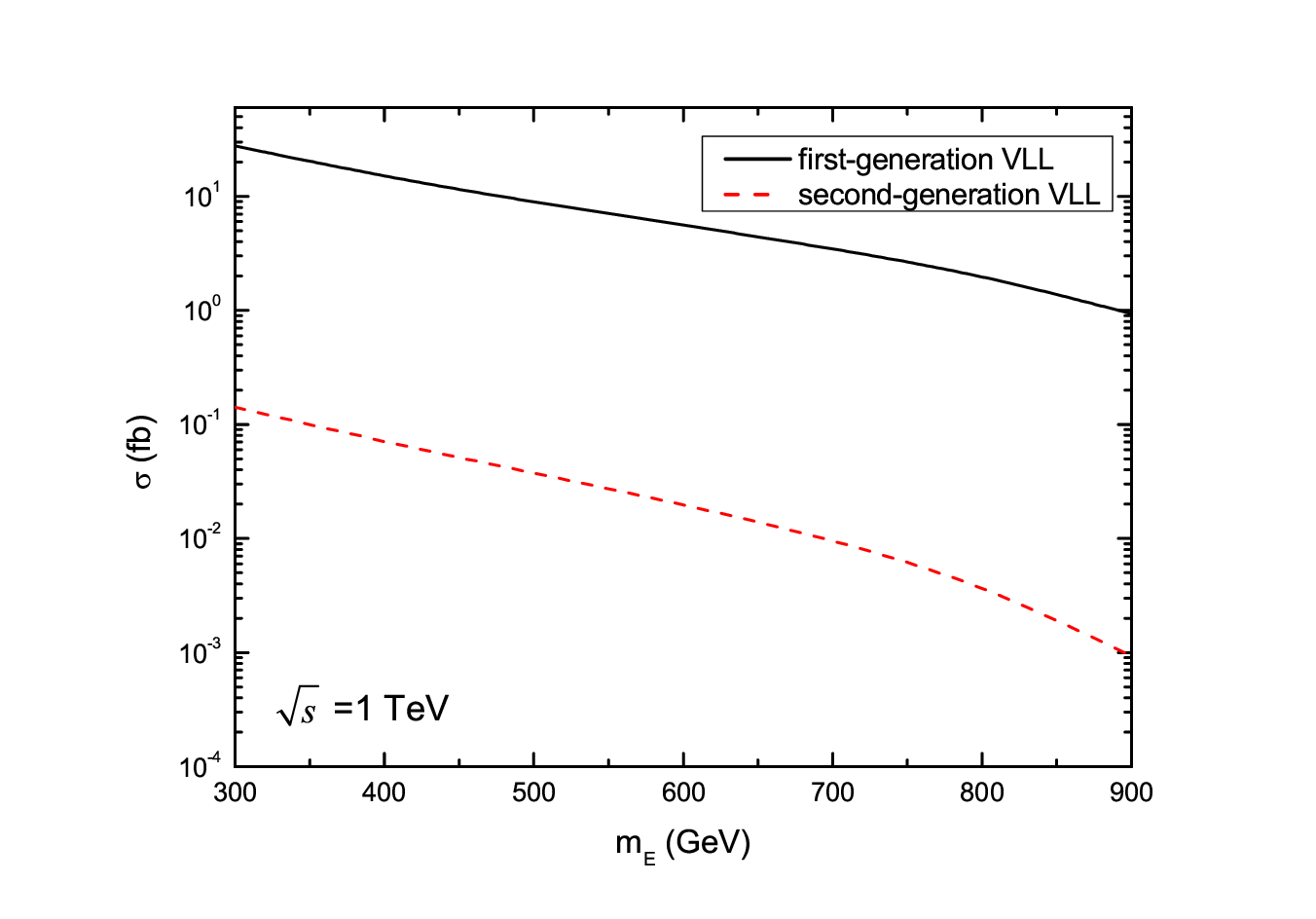}
\hspace{-1.5cm}\epsfxsize=9cm\epsffile{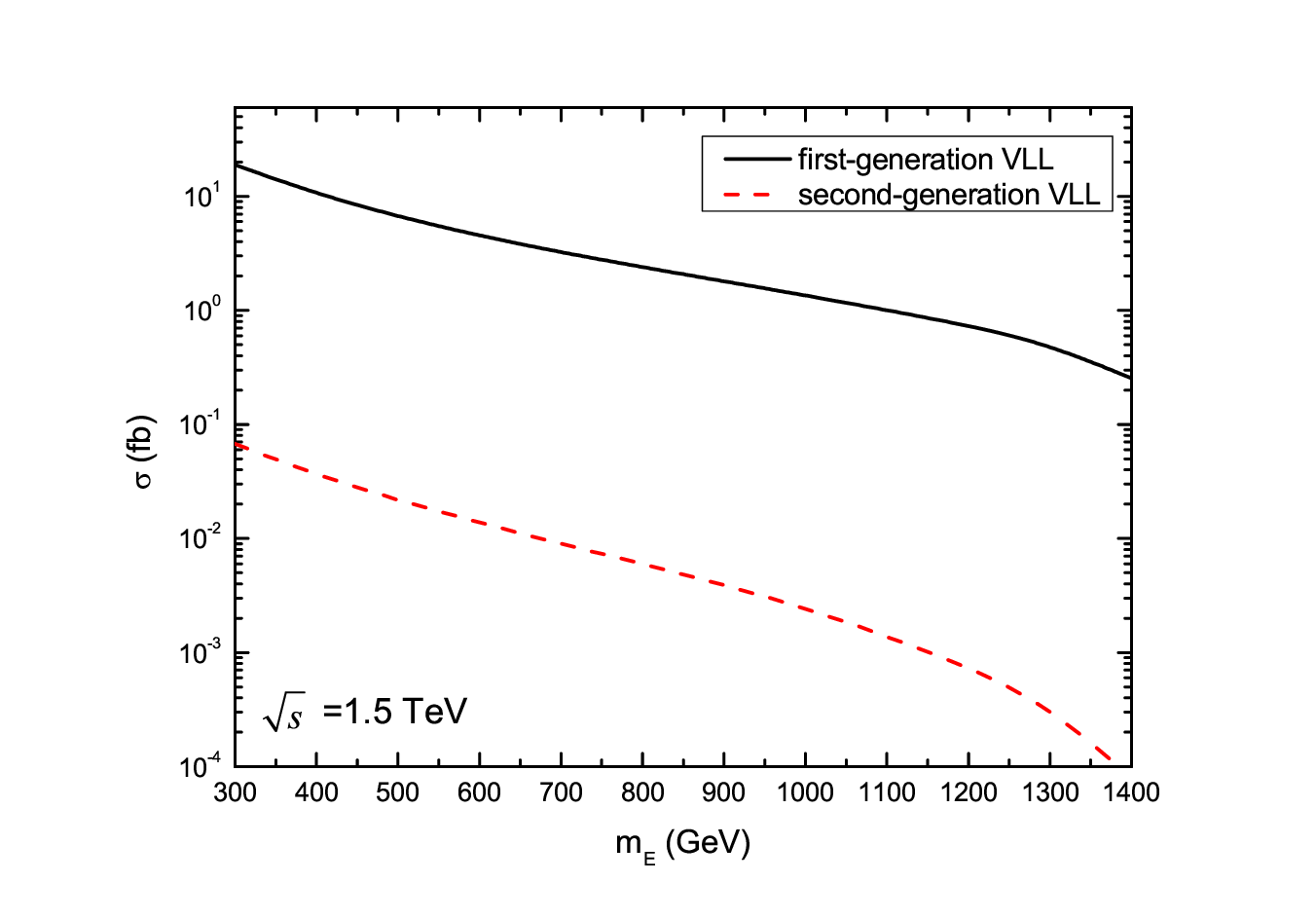}}
\caption{Production cross sections for $e^+ e^- \rightarrow e^\pm E^\mp$ as functions of the VLL mass at $\sqrt{s}=1$ TeV (left) and 1.5 TeV (right), assuming $\epsilon=0.1$.}
\label{fig2}
\end{center}
\end{figure}

Figure~\ref{fig2} shows the cross sections for \(e^+e^- \to e^{\pm}E^{\mp}\) as functions of the VLL mass at \(\sqrt{s}=1\) TeV (left) and 1.5 TeV (right), assuming \(\epsilon = 0.1\), for the singlet first-generation (solid curves) and second-generation (dashed curves) VLLs. The first-generation production receives contributions from both \(s\)- and \(t\)-channel exchanges, with the latter being dominant. This yields cross sections of about 8.9 fb at \(m_E = 500\) GeV and 0.9 fb at 900 GeV for \(\sqrt{s}=1\) TeV, and about 1.36 fb at \(m_E = 1000\) GeV and 0.26 fb at 1400 GeV for \(\sqrt{s}=1.5\) TeV. In contrast, the second-generation VLL, lacking \(t\)-channel couplings, can only be produced via the \(s\)-channel. Notably, the cross sections for second-generation production, amounting to \(4.0\times10^{-2}\) fb and \(9.0\times10^{-4}\) fb at 500 and 900 GeV for \(\sqrt{s}=1\) TeV, and \(2.5\times10^{-3}\) fb and \(8.0\times10^{-4}\) fb at 1000 and 1400 GeV for \(\sqrt{s}=1.5\) TeV, precisely represent the \(s\)-channel contribution that is also present in the first-generation case.

This stark difference underscores the dominance of the \(t\)-channel contribution in first-generation VLL production, which greatly enhances the cross section compared to the pure \(s\)-channel process relevant to second-generation (and similarly third-generation) VLLs. This motivates our focus on the first-generation scenario, where the sizable production rates offer realistic discovery prospects at future \(e^+e^-\) colliders.

We note that the cross sections shown in Fig.~2 are computed without beam polarization. In the collider analysis, we adopt the fixed benchmark beam-polarization
configuration
$P(e^-)=-80\%$ and $P(e^+)=+30\%$,
for all signal and background samples. This configuration enhances the
signal production cross section by approximately a factor of 1.6 for
the collider energies considered.

\section{Collider Analysis and Signal Sensitivity}
\label{sec:analysis}

\subsection{Signal Signatures and Numerical Simulations}

We evaluate the discovery potential via detector-level Monte Carlo (MC) simulations of both signal and SM background processes. For heavy VLL masses, the $W$ and $Z$ bosons produced in the decays are highly boosted, such that their hadronic decay products merge into a single fat-jet, denoted as $J$. This fat-jet topology provides a powerful handle for signal identification and background suppression. For the Higgs channel, the fat-jet approach is applicable at $\sqrt{s}=1.5\ \text{TeV}$, where the Higgs is sufficiently boosted; at $\sqrt{s}=1\ \text{TeV}$, however, the Higgs decay products are reconstructed as two resolved $b$-jets.

The analysis is based on the three dominant decay modes of the singlet vector-like electron:
$E^{\pm}\to W^{\pm}\nu_{e}$, $E^{\pm}\to Z e^{\pm}$ and $E^{\pm}\to h e^{\pm}$, which in the high-mass limit have branching ratios in the proportion $2:1:1$. Considering the hadronic decays of the gauge bosons, $W^{\pm}\to q\bar{q}'$, $Z\to q\bar{q}$ and $h\to b\bar{b}$, we consider the following three final-state topologies:

\textbf{Case 1 ($W$ channel):}
\begin{equation}
e^{+} e^{-} \to e^{ \pm} E^{\mp} \to e^{ \pm}+J+\not{E}_T,
\label{eq:case1}
\end{equation}
with $E^{\pm}\to W^{\pm}\nu_{e}$, $W^{\pm}\to q\bar{q}'$.

\textbf{Case 2 ($Z$ channel):}
\begin{equation}
e^{+}e^{-}\to e^{\pm } E^{\mp }\to e^{+} e^{-}+J,
\label{eq:case2}
\end{equation}
with $E^{\pm}\to e^{\pm}Z$, $Z\to q\bar{q}$.

\textbf{Case 3 ($h$ channel):}
\begin{equation}
e^+e^-\to e^\pm E^\mp
\to e^+e^- h
\to e^+e^- b\bar b ,
\label{eq:case3}
\end{equation}
Here, the $b\bar b$ system from the Higgs decay is reconstructed as
two resolved $b$-tagged jets at $\sqrt{s}=1~{\rm TeV}$, whereas at
$\sqrt{s}=1.5~{\rm TeV}$ it is reconstructed as a single boosted
Higgs fat jet.

To test the scope of future $e^+e^-$ colliders in probing our VLL scenario, we choose four benchmark configurations, with the $E^{\pm}$ mass spanning values from 800 to 1400 GeV:
\begin{itemize}
\item $m_E = 800$ and 900 GeV at $\sqrt{s} = 1$ TeV (ILC);
\item $m_E = 1300$ and 1400 GeV at $\sqrt{s} = 1.5$ TeV (CLIC).
\end{itemize}

Signal and SM background events are generated at leading order (LO) using MadGraph5\_aMC@NLO~\cite{Alwall:2014hca}, followed by parton showering and hadronization with Pythia8.20~\cite{Sjostrand:2014zea}. Detector effects are simulated using Delphes 3.4.2~\cite{deFavereau:2013fsa} with the ILD detector card~\cite{Behnke:2013lya,ILDConceptGroup:2020sfq}. For the identification of $b$-jets, we adopt the default $b$-tagging settings of the ILD card~\cite{Behnke:2013lya} (implemented as \texttt{Delphes\_card\_ILD.tcl} in Delphes)~\footnote{For a uniform detector-level comparison, the ILD Delphes
parameterization is used for both collider energies. We therefore do
not attempt a CLIC-specific detector-performance study in the present
analysis, and the 1.5~TeV results should be interpreted within this
common detector benchmark.}, where the tagging efficiencies depend on the jet transverse momentum $p_T$ and are based on the parameterization of Ref.~\cite{CMS:2012feb}. For a typical $b$-jet with $p_T = 100$~GeV, this corresponds to a $b$-tagging efficiency of approximately $78\%$, with $c$-jet and light-jet mis-tag rates of about $21\%$ and $1.4\%$, respectively. For the reconstruction of fat-jets, the Cambridge-Aachen algorithm~\cite{Dokshitzer:1997in,Wobisch:1998wt}, as implemented in the FastJet package~\cite{Cacciari:2011ma}, is employed with a cone radius parameter of $R = 1.0$. The event analysis is carried out within the MadAnalysis5 framework~\cite{ma5,Conte:2014zja}.

The following basic kinematic requirements are imposed on the generated leptons and jets:
\begin{equation}
p_T^\ell > 10\ \text{GeV},\quad |\eta_\ell| < 2.5,\quad p_T^j > 20\ \text{GeV},\quad |\eta_j| < 5,
\end{equation}
where $\ell$ and $j$ denote leptons and jets, respectively.

Initial-state radiation (ISR) and beamstrahlung effects are not
included in the present event simulation. Accordingly, both the
production rates and the event kinematics used in our analysis are
evaluated at the nominal center-of-mass energies,
$\sqrt{s}=1~{\rm TeV}$ and $1.5~{\rm TeV}$. These effects can in
principle modify the total production rates and broaden the
kinematic distributions, in particular the associated-electron
energy and the corresponding recoil-mass structure. A dedicated
treatment of ISR, beamstrahlung, and the realistic luminosity
spectrum is beyond the scope of the present baseline cut-based
analysis. The sensitivities reported below should therefore be
interpreted within this fixed-$\sqrt{s}$ approximation.
\subsection{Recoil-mass relation and common associated-electron selection}
\label{sec:recoil}

A common feature of all three signal channels is the electron produced
in association with the VLL in the single-production process
\begin{equation}
e^+e^- \to e^\pm E^\mp .
\end{equation}
Its kinematics are determined primarily by the VLL mass and the
center-of-mass energy, and are therefore largely independent of the
subsequent decay mode of the heavy VLL. We consequently use the
associated-electron energy as a common mass-sensitive observable in
all three channels.

The recoil mass against the associated electron is defined as
\begin{equation}
M_{\mathrm{recoil}}^2 =
\left(
p_{e^+}+p_{e^-}-p_{e_{\rm assoc}}
\right)^2 ,
\end{equation}
where $p_{e_{\rm assoc}}$ denotes the four-momentum of the electron
produced in association with the VLL. In the center-of-mass frame,
neglecting the electron mass and using the nominal collision energy
$\sqrt{s}$, this relation becomes
\begin{equation}
M_{\mathrm{recoil}}^2
\simeq
s-2\sqrt{s}\,E_{e_{\rm assoc}} .
\label{eq:recoilmass}
\end{equation}
Here and throughout the present kinematic analysis, $\sqrt{s}$ denotes
the nominal center-of-mass energy. ISR and beamstrahlung effects are
not included in the present simulation. Thus, within the fixed-$\sqrt{s}$
approximation adopted here, the associated-electron energy
$E_{e_{\rm assoc}}$ is in one-to-one correspondence with the recoil
mass. For the two-body production process, its characteristic energy
is
\begin{equation}
E_{e_{\rm assoc}}^{\mathrm{peak}}
\simeq
\frac{s-m_E^2}{2\sqrt{s}},
\label{eq:associateenergypeak}
\end{equation}
which corresponds to
$M_{\mathrm{recoil}}\simeq m_E$.

Motivated by the narrow signal structure in the associated-electron
energy distribution, we define a mass-hypothesis-dependent condition
corresponding to
\begin{equation}
\left|M_{\mathrm{recoil}}-m_E\right|
<20~{\rm GeV}.
\label{eq:recoilwindow}
\end{equation}
In the present analysis, $M_{\rm recoil}$ is not introduced as an
independent event-selection variable. Instead, Eq.~(\ref{eq:recoilmass}) allows the
recoil-mass information to be implemented directly through the
associated-electron energy $E_{e_{\rm assoc}}$. Since the two
quantities are in one-to-one correspondence at fixed nominal
$\sqrt{s}$, a mass-hypothesis-dependent selection on
$E_{e_{\rm assoc}}$ is kinematically equivalent to selecting a
recoil-mass window centered on the assumed VLL mass. In particular,
the characteristic signal energies given by Eq.~(\ref{eq:associateenergypeak}) correspond to
$M_{\rm recoil}\simeq m_E$ for the respective mass hypotheses.

A common half-width of $20~{\rm GeV}$ is adopted as a benchmark choice
for all mass hypotheses and all three signal channels. Within our
detector-level simulation, this choice retains the narrow signal region
while providing effective suppression of the continuum background.
The window is kept fixed rather than optimized point by point, in
accordance with the baseline character of the present cut-based
analysis. A dedicated collider-specific analysis including ISR, beamstrahlung,
the realistic luminosity spectrum, and a corresponding optimization
of the recoil-mass window is left for future work.

  Using
Eq.~(\ref{eq:recoilmass}), the recoil-mass condition in
Eq.~(\ref{eq:recoilwindow}) is translated into the associated-electron
energy window
\begin{equation}
\frac{s-(m_E+20~{\rm GeV})^2}{2\sqrt{s}}
<
E_{e_{\rm assoc}}
<
\frac{s-(m_E-20~{\rm GeV})^2}{2\sqrt{s}} .
\label{eq:associateenergywindow}
\end{equation}

For the benchmark masses considered here, the corresponding
associated-electron energy windows at $\sqrt{s}=1~{\rm TeV}$ are
\[
163.8<E_{e_{\rm assoc}}<195.8~{\rm GeV},
\qquad
76.8<E_{e_{\rm assoc}}<112.8~{\rm GeV},
\]
for $m_E=800$ and $900~{\rm GeV}$, respectively. At
$\sqrt{s}=1.5~{\rm TeV}$, they become
\[
169.2<E_{e_{\rm assoc}}<203.9~{\rm GeV},
\qquad
77.9<E_{e_{\rm assoc}}<115.2~{\rm GeV},
\]
for $m_E=1300$ and $1400~{\rm GeV}$, respectively.

In Case~1, the single reconstructed electron is directly identified
as the associated electron. In Cases~2 and 3, an additional electron arises from the VLL decay.
For the benchmark masses considered, the electron from the VLL decay
is predominantly the harder one, whereas the electron produced at the
single-production vertex is typically the sub-leading one.
Accordingly, the sub-leading electron is taken as the
associated-electron candidate and is used to implement the
mass-hypothesis-dependent $E_{e_{\rm assoc}}$ selection defined above.

\subsection{Case 1: $e^{\pm}+J+\not{E}_T$}

In addition to the common mass-hypothesis-dependent
associated-electron energy selection described in
Sec.~\ref{sec:recoil}, we impose requirements specific to the
$W\nu$ channel. The dominant SM background arises from
$e^+e^- \to e^{\pm}\nu jj$,
which receives contributions from $WW$ production, $t$-channel
processes, and off-shell gauge-boson exchanges. In these background
events, the hadronic system can be reconstructed as a fat-jet $J$,
while the isolated electron and missing transverse momentum mimic the
signal topology. The signal, in contrast, originates from
$e^+e^-\to e^\pm E^\mp$, followed by
$E^\pm\to W^\pm\nu_e$, with the boosted hadronically decaying
$W$ boson reconstructed as a single fat-jet.

\begin{figure}[ht]
\begin{center}
\centerline{\hspace{2.0cm}\epsfxsize=8cm\epsffile{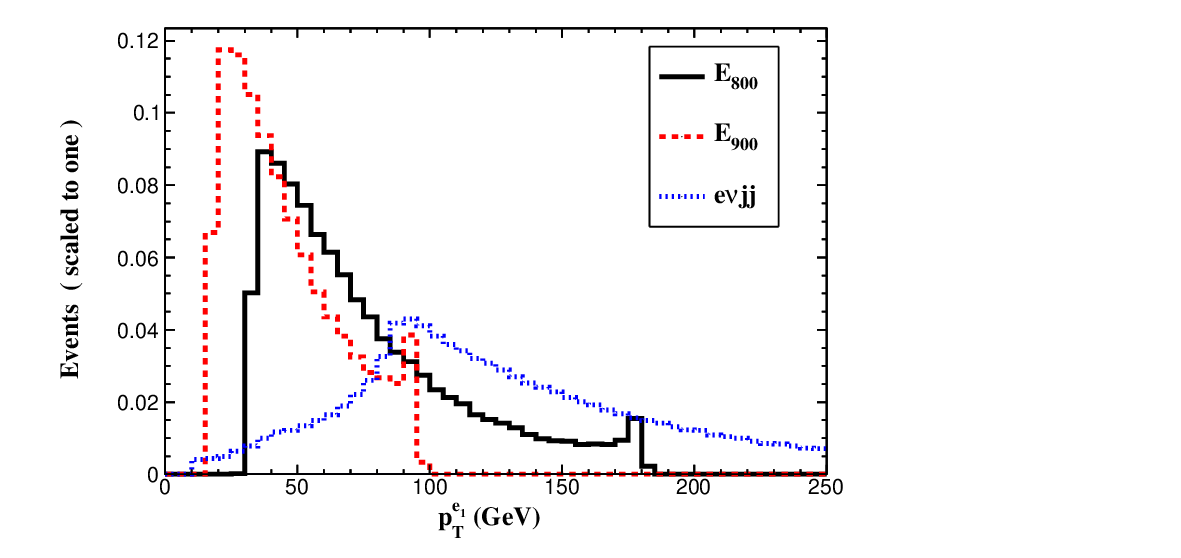}
\hspace{-2.0cm}\epsfxsize=8cm\epsffile{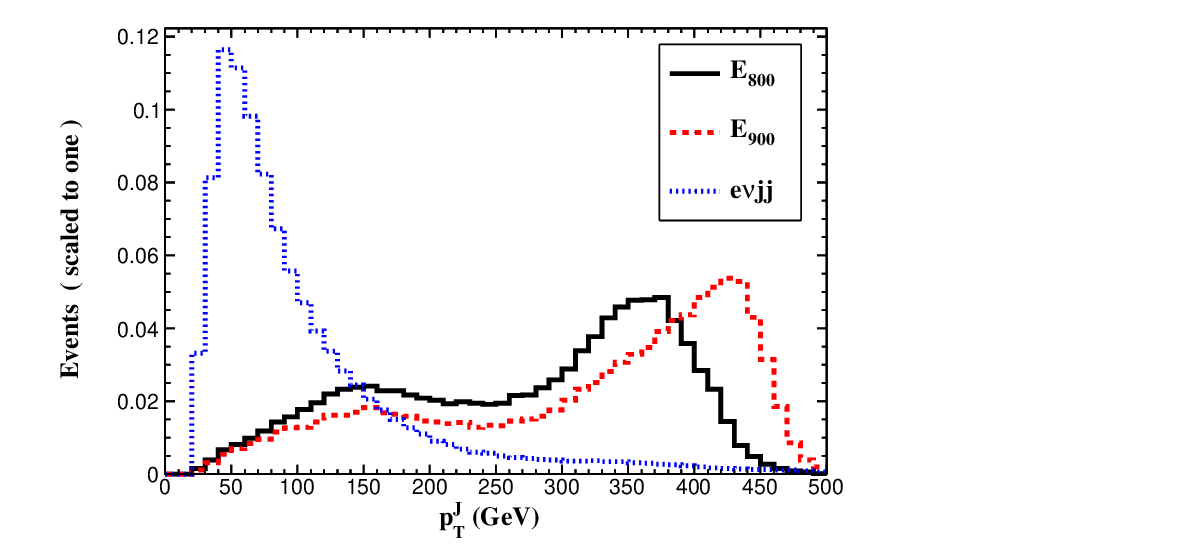}}
\centerline{\hspace{2.0cm}\epsfxsize=8cm\epsffile{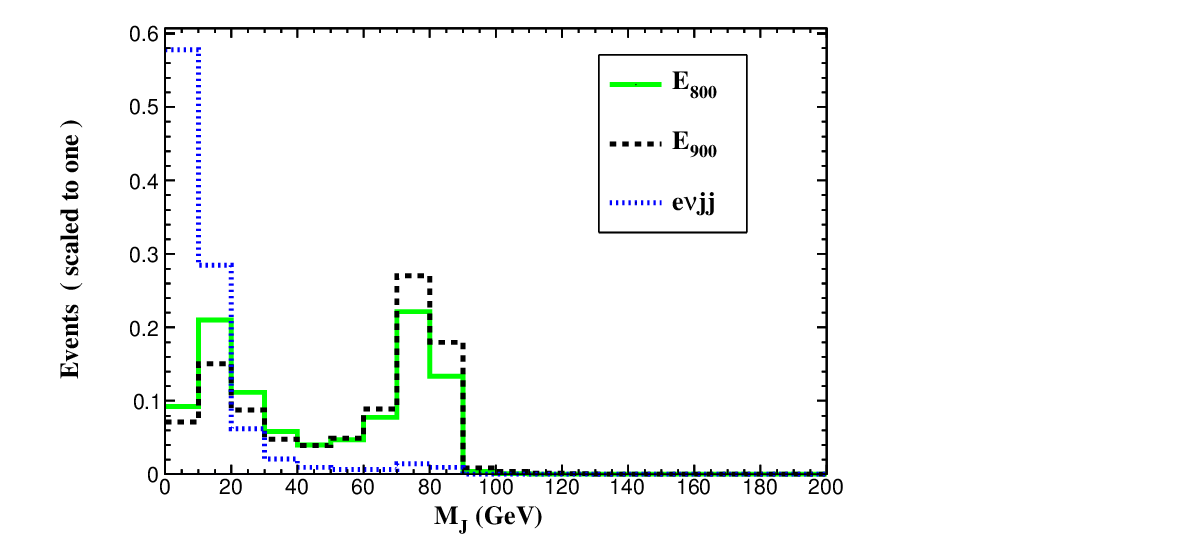}
\hspace{-2.0cm}\epsfxsize=8cm\epsffile{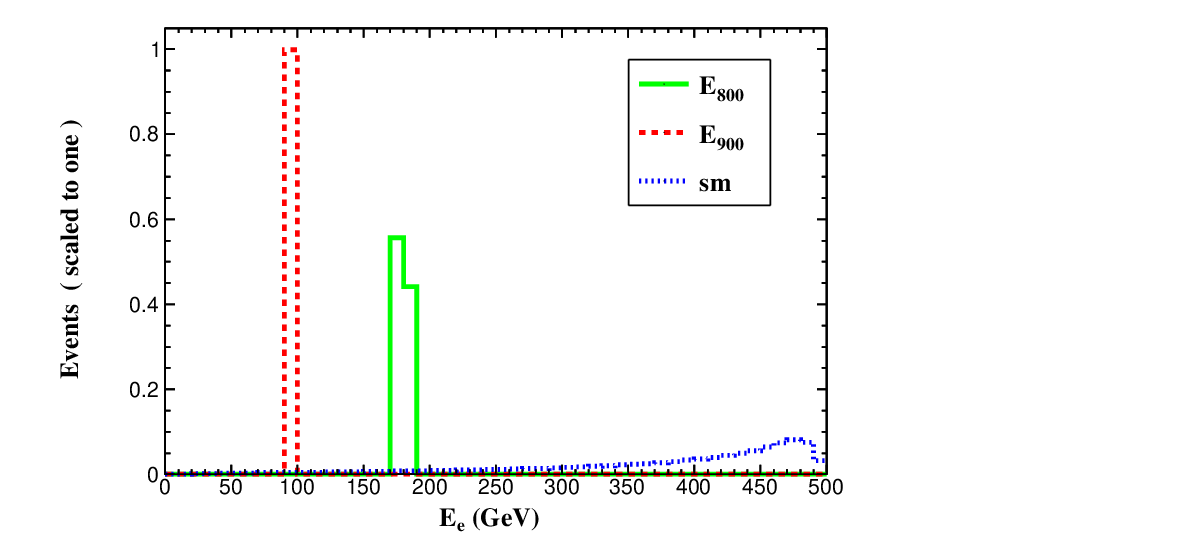}}
\caption{Normalized distributions of the main kinematic observables
for Case~1 and the dominant SM background at $\sqrt{s}=1$~TeV,
for $m_E=800$ and $900$~GeV.}
\label{fig3}
\end{center}
\end{figure}

\begin{figure}[ht]
\begin{center}
\centerline{\hspace{2.0cm}\epsfxsize=8cm\epsffile{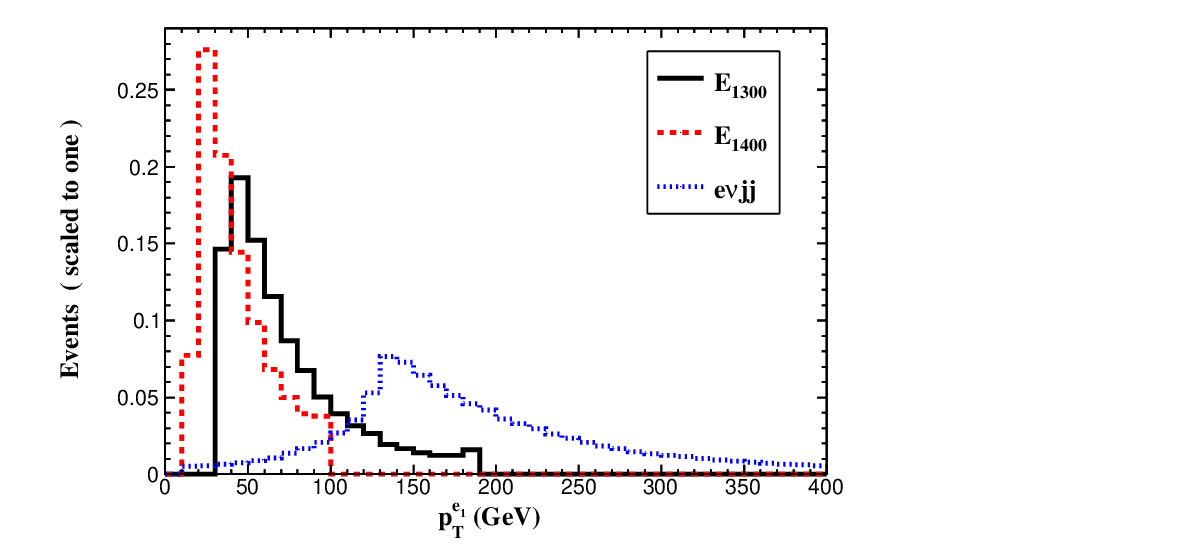}
\hspace{-2.0cm}\epsfxsize=8cm\epsffile{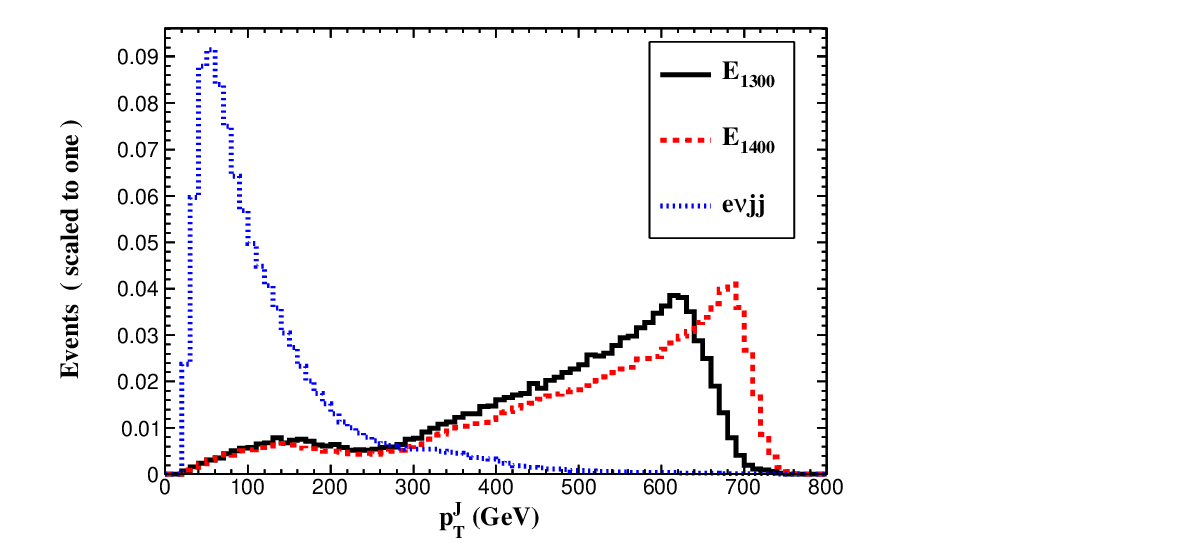}}
\centerline{\hspace{2.0cm}\epsfxsize=8cm\epsffile{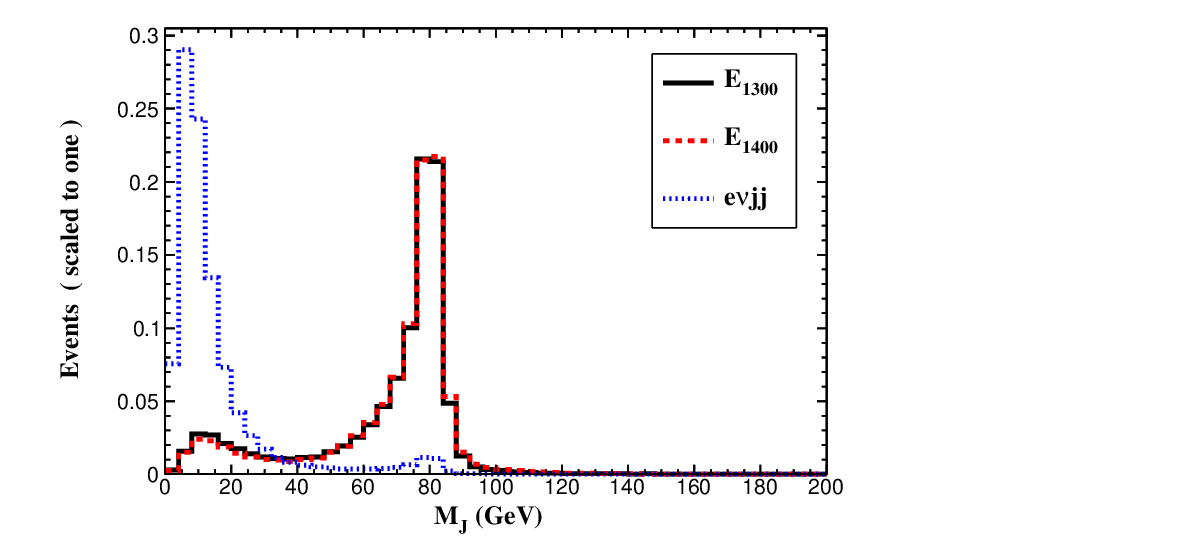}
\hspace{-2.0cm}\epsfxsize=8cm\epsffile{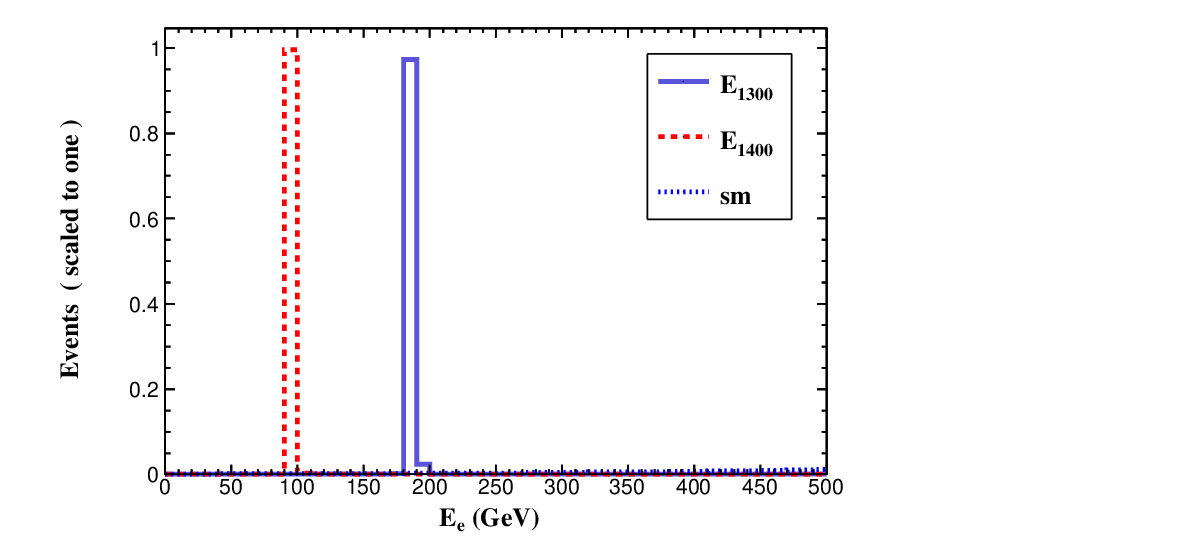}}
\caption{Same as Fig.~\ref{fig3}, but for $\sqrt{s}=1.5$~TeV with
$m_E=1300$ and $1400$~GeV.}
\label{fig4}
\end{center}
\end{figure}

The normalized distributions of the main kinematic observables for
the signal benchmarks and the dominant SM background are shown in
Figs.~\ref{fig3} and~\ref{fig4}. We consider the transverse momentum
of the electron, $p_T^e$, the transverse momentum and invariant mass
of the fat-jet, $p_T^J$ and $M_J$, and the energy of the associated
electron, $E_e$. In particular, $E_e$ provides a direct
mass-sensitive handle through the recoil-mass relation discussed in
Sec.~\ref{sec:recoil}. For fixed nominal $\sqrt{s}$,
$E_e$ and $M_{\rm recoil}$ are in one-to-one correspondence through
Eq.~(\ref{eq:recoilmass}). We therefore use $E_e$ directly as the practical event-selection
variable, since it contains the same mass-sensitive information as
$M_{\rm recoil}$ within the fixed-$\sqrt{s}$ approximation.

The event selection consists of three sequential requirements. Since only
one isolated electron is present in this channel, it is directly identified
as the associated electron, allowing the mass-sensitive energy selection to
be applied first.
\begin{itemize}

\item \textbf{Cut 1:}
Exactly one isolated electron is required, with its energy satisfying the
mass-hypothesis-dependent window defined in
Eq.~(\ref{eq:associateenergywindow}), corresponding to
$|M_{\rm recoil}-m_E|<20~{\rm GeV}$.

\item \textbf{Cut 2:}
At least one fat-jet $J$ is required, with
$p_T^J>100$~GeV for $\sqrt{s}=1$~TeV and
$p_T^J>200$~GeV for $\sqrt{s}=1.5$~TeV, together with
$|M_J-m_W|<25~{\rm GeV}$.

\item \textbf{Cut 3:}
The transverse momentum of the associated electron is required to satisfy
$p_T^e<100$~GeV.

\end{itemize}

\begin{table}[!h]
\centering
\small
\setlength{\tabcolsep}{4pt}
\caption{Cross sections (in fb) for the signal benchmarks and the SM
background in Case~1 at $\sqrt{s}=1$ and $1.5$~TeV, assuming
$\epsilon=0.1$. For Cuts~1--3, the unbracketed background entries
correspond to the $m_E=800$ and $1300$~GeV hypotheses at
$\sqrt{s}=1$ and $1.5$~TeV, respectively, while the entries in
square brackets correspond to the $m_E=900$ and $1400$~GeV
hypotheses.}
\label{cutflow1}
\vspace{0.1cm}
\begin{tabular}{l ccc ccc}
\toprule[1.5pt]
& \multicolumn{3}{c}{$\sqrt{s}=1$ TeV}
& \multicolumn{3}{c}{$\sqrt{s}=1.5$ TeV} \\
\cmidrule{2-4} \cmidrule{5-7}
\multirow{2}{*}{Cuts}
& \multicolumn{2}{c}{Signal} & Background
& \multicolumn{2}{c}{Signal} & Background \\
& 800~GeV & 900~GeV & $e^{\pm}\nu jj$
& 1300~GeV & 1400~GeV & $e^{\pm}\nu jj$ \\
\midrule[0.8pt]
Basic & 0.87 & 0.40 & 1502 & 0.21 & 0.11 & 1258 \\
Cut 1 & 0.87 & 0.40 & 36.2~[21.7] & 0.21 & 0.11 & 11.0~[7.9] \\
Cut 2 & 0.42 & 0.24 & 2.39~[2.12] & 0.167 & 0.088 & 1.1~[0.78] \\
Cut 3 & 0.33 & 0.24 & 0.66~[1.95] & 0.137 & 0.088 & 0.67~[0.74] \\
\bottomrule[1.5pt]
\end{tabular}
\end{table}

The resulting cut flow is summarized in
Table~\ref{cutflow1}. The mass-dependent
associated-electron energy requirement
(Cut~1) is imposed first and provides the
dominant initial suppression of the SM
background while leaving the signal
essentially unchanged. The subsequent
fat-jet selection (Cut~2), including the
transverse-momentum requirement and the
$W$-candidate mass window, together with
the electron-$p_T$ requirement (Cut~3),
further reduces the residual background.
The selection thresholds are chosen to
retain signal acceptance while effectively
suppressing the continuum background.
After all selections, the background cross
sections are $0.66$~fb and $1.95$~fb for
$m_E=800$ and $900$~GeV at
$\sqrt{s}=1$~TeV, and $0.67$~fb and
$0.74$~fb for $m_E=1300$ and $1400$~GeV at
$\sqrt{s}=1.5$~TeV, respectively.

\subsection{Case 2: $e^+e^-+J$}
The signal process is
\(e^+e^-\to e^{\pm}E^{\mp}\) with \(E^{\pm}\to e^{\pm}Z\), where the
hadronically decaying boosted \(Z\) boson is reconstructed as a single fat
jet \(J\).
The dominant SM background arises from
$e^+e^-\to e^+e^-jj$,
which receives contributions from $ZZ$ production,
off-shell $Z/\gamma^*$ exchange, and $t$-channel diagrams. These
processes can produce two isolated electrons together with a hadronic
system that satisfies the fat-jet requirements and thereby mimics the
signal topology.

\begin{figure*}[ht]
\begin{center}
\centerline{\hspace{2.0cm}\epsfxsize=9cm\epsffile{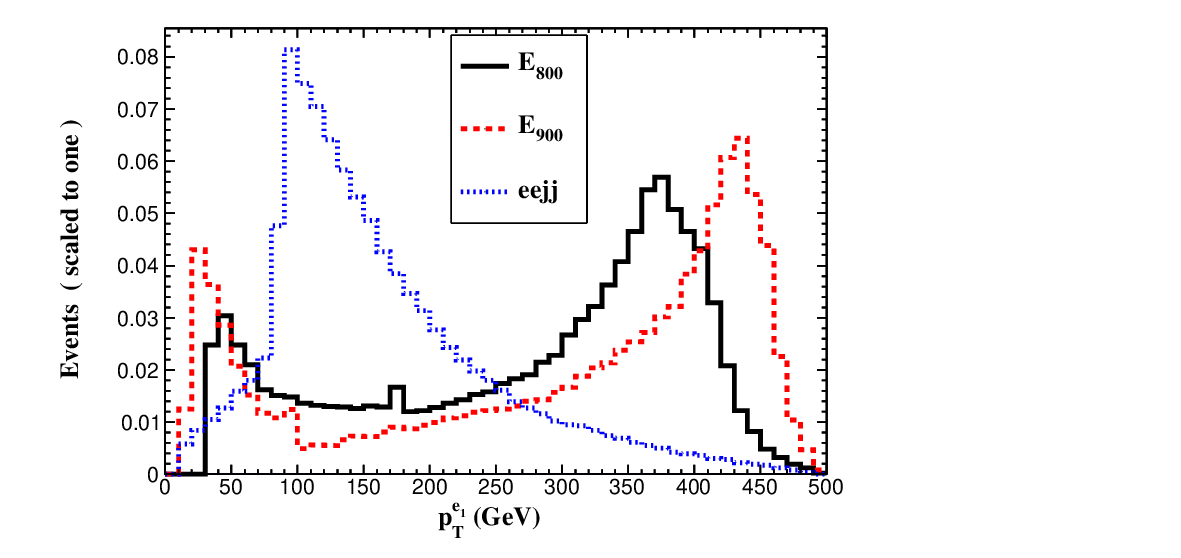}
\hspace{-2.0cm}\epsfxsize=9cm\epsffile{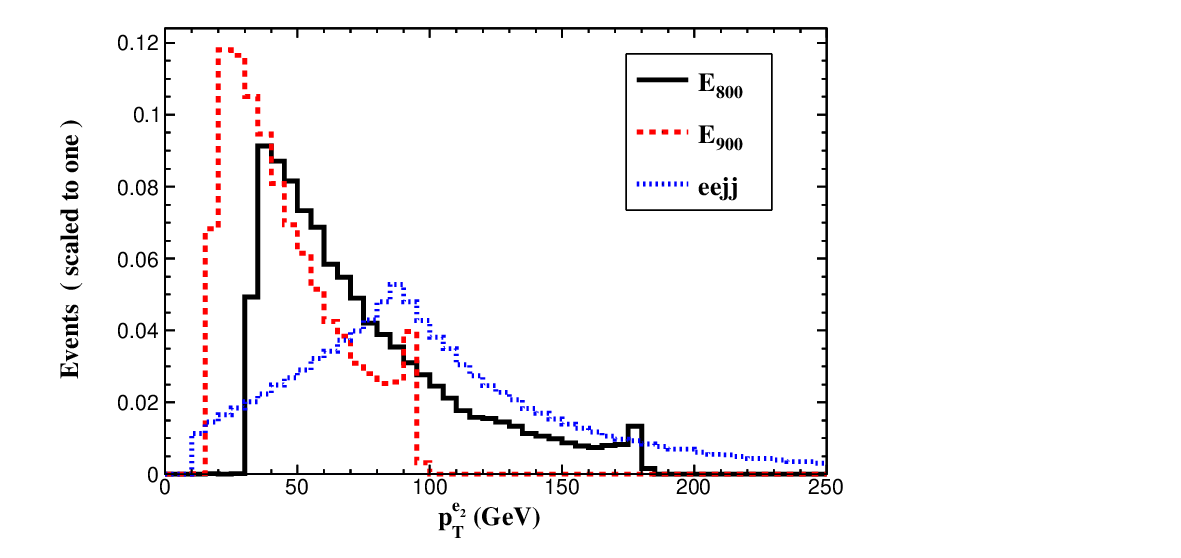}}
\centerline{\hspace{2.0cm}\epsfxsize=9cm\epsffile{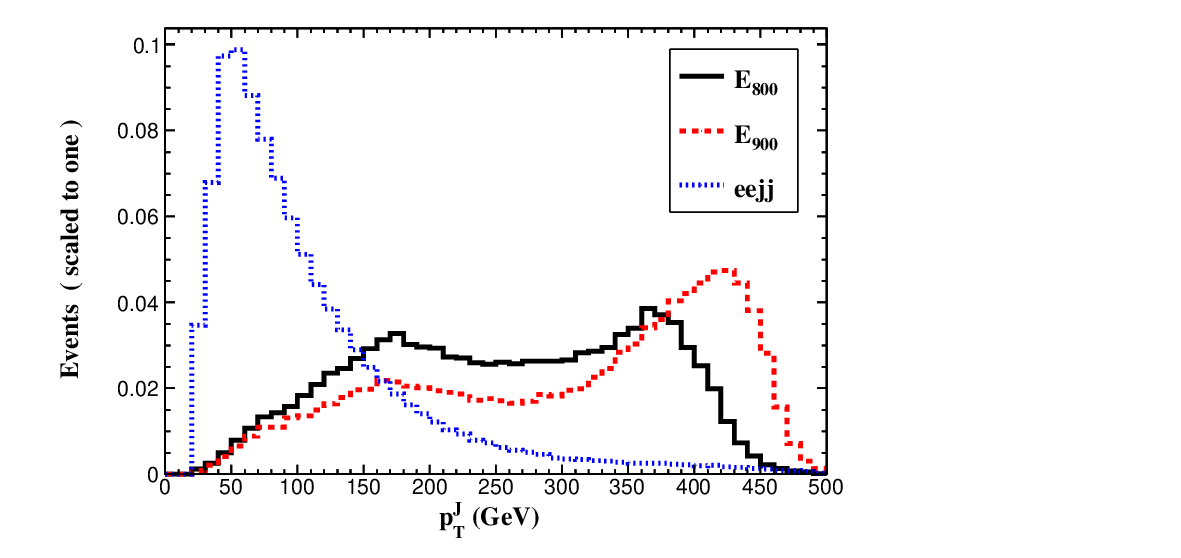}
\hspace{-2.0cm}\epsfxsize=9cm\epsffile{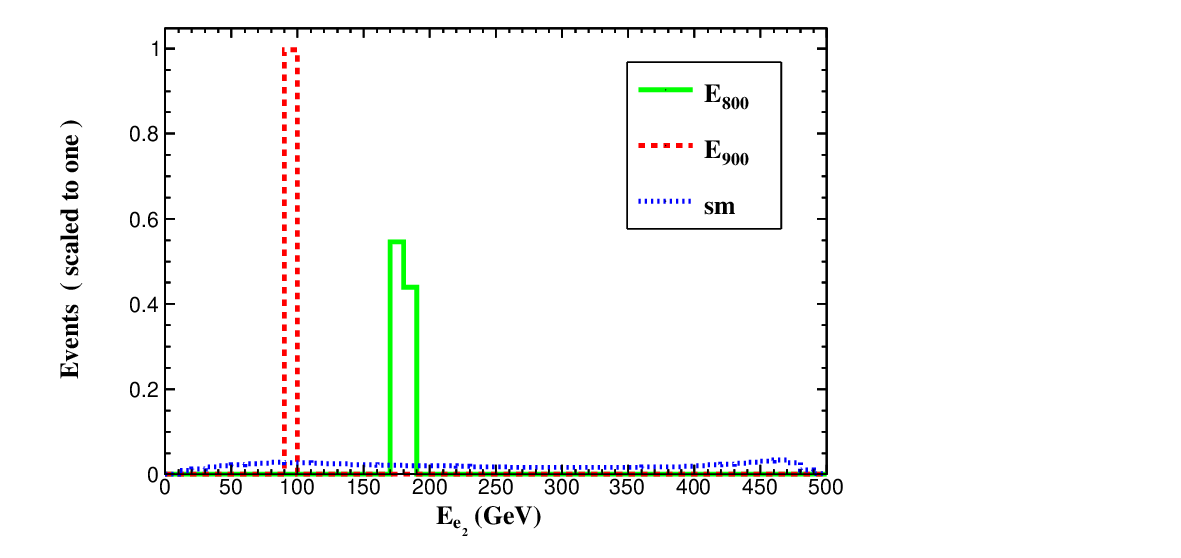}}
\caption{Normalized distributions of the main kinematic observables
for Case~2 and the dominant SM background at $\sqrt{s}=1$~TeV,
for $m_E=800$ and $900$~GeV.}
\label{fig5}
\end{center}
\end{figure*}

\begin{figure*}[ht]
\begin{center}
\centerline{\hspace{2.0cm}\epsfxsize=9cm\epsffile{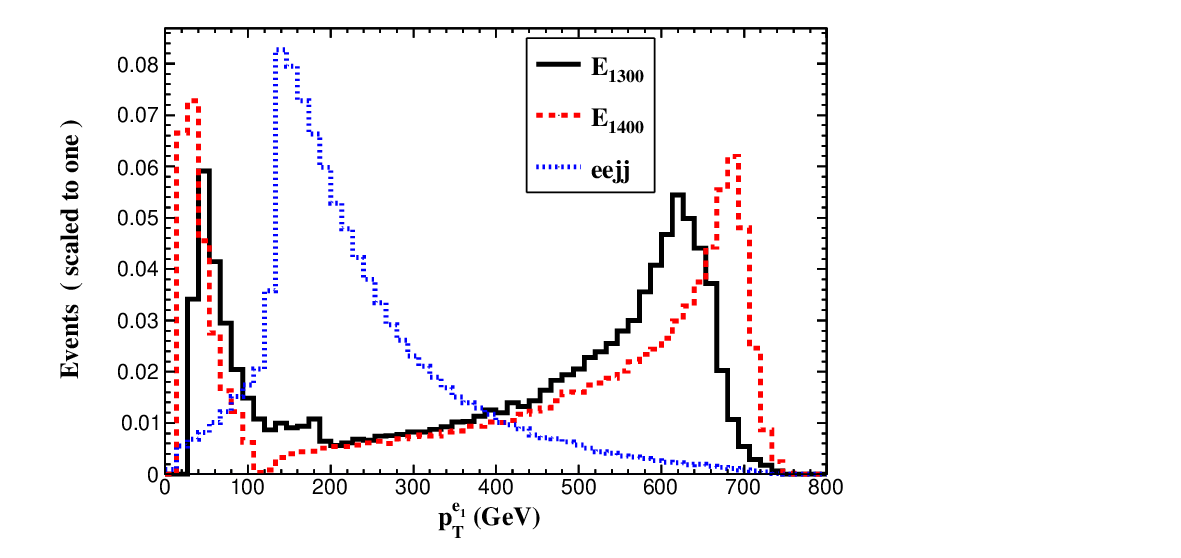}
\hspace{-2.0cm}\epsfxsize=9cm\epsffile{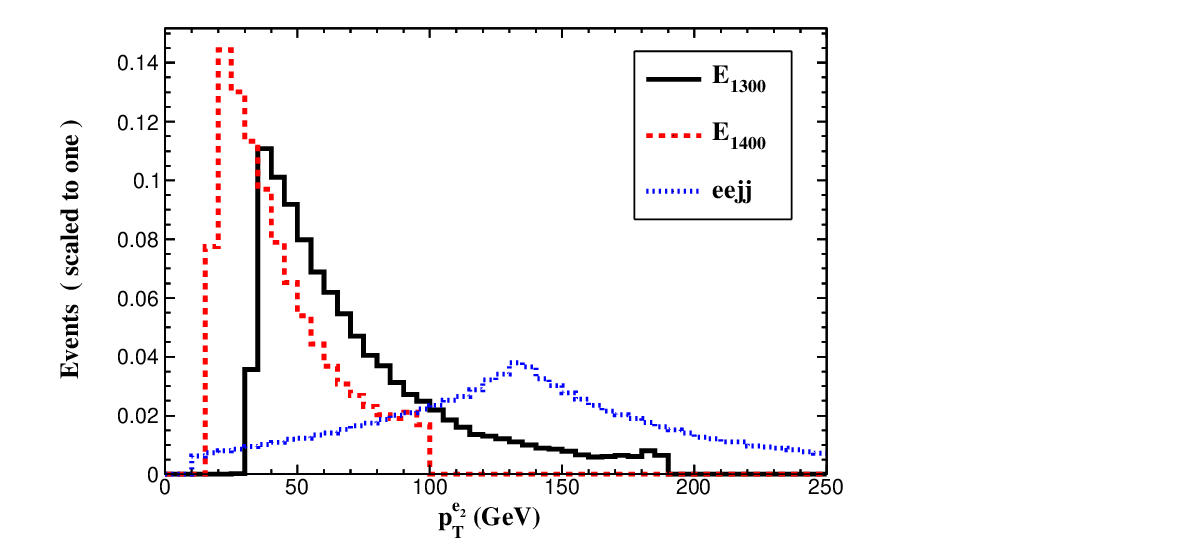}}
\centerline{\hspace{2.0cm}\epsfxsize=9cm\epsffile{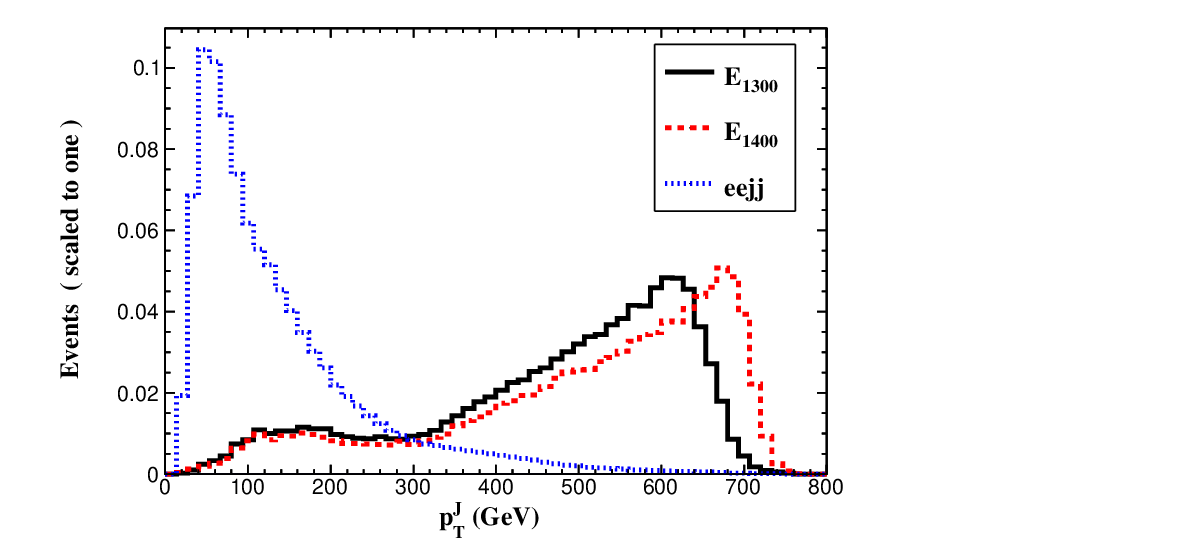}
\hspace{-2.0cm}\epsfxsize=9cm\epsffile{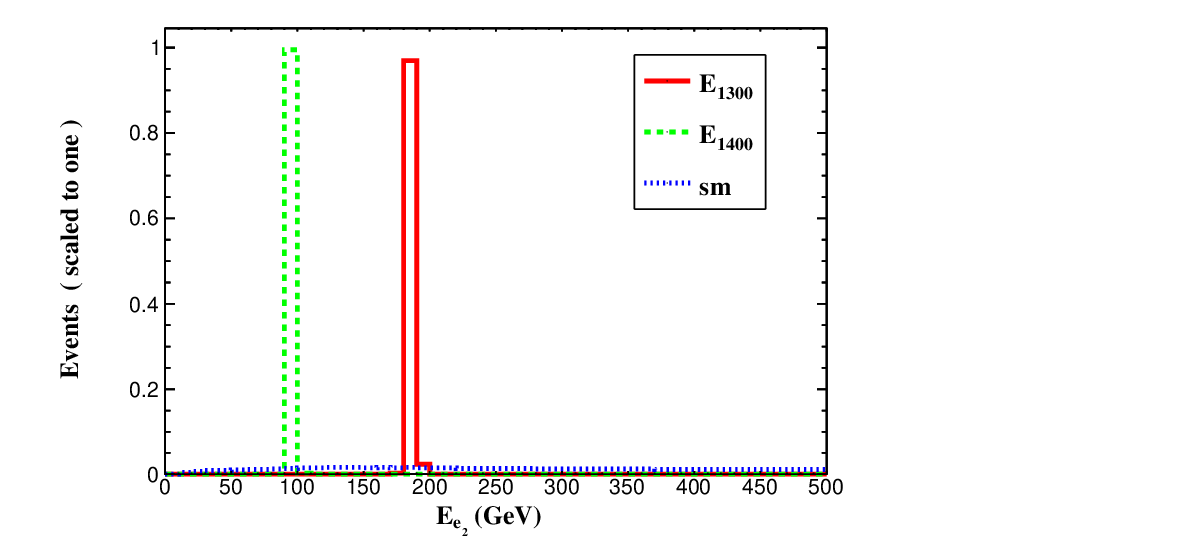}}
\caption{Same as Fig.~\ref{fig5}, but for $\sqrt{s}=1.5$~TeV with
$m_E=1300$ and $1400$~GeV.}
\label{fig6}
\end{center}
\end{figure*}

For the benchmark masses considered, the electron from the heavy-VLL decay is
typically much harder than the electron produced in association with the VLL. We denote the leading and
sub-leading electrons by $e_1$ and $e_2$, respectively, and use
$e_2$ as the associated-electron candidate in the
mass-hypothesis-dependent selection.
The normalized distributions of the main kinematic observables for the
signal benchmarks and the dominant SM background are shown in
Figs.~\ref{fig5} and~\ref{fig6} for $\sqrt{s}=1$ and $1.5$~TeV,
respectively. We consider the transverse momenta of the leading and
sub-leading electrons, $p_T^{e_1}$ and $p_T^{e_2}$, the fat-jet
transverse momentum $p_T^J$, and the energy of the associated electron
$E_{e_2}$. As discussed in Sec.~\ref{sec:recoil}, $E_{e_2}$ is
directly related to the VLL recoil mass and therefore provides a
mass-sensitive observable for the event selection.

The event selection consists of three sequential requirements. The
two-electron topology is first used to identify the associated-electron
candidate, after which the mass-sensitive energy selection is applied
before the hadronic $Z$ reconstruction.

\begin{itemize}

\item \textbf{Cut 1:}
Exactly two isolated, oppositely charged electrons are required.
The leading electron must satisfy
$p_T^{e_1}>250$~GeV for $\sqrt{s}=1$~TeV and
$p_T^{e_1}>400$~GeV for $\sqrt{s}=1.5$~TeV, while the sub-leading
electron is required to satisfy $p_T^{e_2}<100$~GeV.
The two electrons are further required to satisfy
$\Delta R_{e_1,e_2}>1.0$.
For the signal events, the leading electron predominantly originates
from the VLL decay, whereas the sub-leading electron is identified
as the associated-electron candidate and is used in the subsequent
mass-hypothesis-dependent energy selection.

\item \textbf{Cut 2:}
The associated-electron energy $E_{e_2}$
is required to satisfy the mass-hypothesis-dependent window
defined in Eq.~(\ref{eq:associateenergywindow}), corresponding to
$|M_{\rm recoil}-m_E|<20$~GeV.

\item \textbf{Cut 3:}
At least one fat-jet $J$ is required, with
$p_T^J>100$~GeV for $\sqrt{s}=1$~TeV and
$p_T^J>200$~GeV for $\sqrt{s}=1.5$~TeV.
Furthermore, the fat-jet invariant mass must be consistent
with the $Z$-boson mass, satisfying
$|M_J-m_Z|<25$~GeV.

\end{itemize}

\begin{table}[h]
\centering
\small
\setlength{\tabcolsep}{4pt}
\caption{Cross sections (in fb) for the signal benchmarks and the SM
background in Case~2 at $\sqrt{s}=1$ and $1.5$~TeV, assuming
$\epsilon=0.1$. For Cuts~2 and~3, the unbracketed background entries
correspond to the $m_E=800$ and $1300$~GeV hypotheses at
$\sqrt{s}=1$ and $1.5$~TeV, respectively, while the entries in
square brackets correspond to the $m_E=900$ and $1400$~GeV
hypotheses.}
\label{cutflow2}
\vspace{0.1cm}
\begin{tabular}{l ccc ccc}
\toprule[1.5pt]
& \multicolumn{3}{c}{$\sqrt{s}=1$ TeV}
& \multicolumn{3}{c}{$\sqrt{s}=1.5$ TeV} \\
\cmidrule{2-4} \cmidrule{5-7}
\multirow{2}{*}{Cuts}
& \multicolumn{2}{c}{Signal} & Background
& \multicolumn{2}{c}{Signal} & Background \\
& 800~GeV & 900~GeV & $e^{+}e^{-}jj$
& 1300~GeV & 1400~GeV & $e^{+}e^{-}jj$ \\
\midrule[0.8pt]
Basic & 0.51 & 0.23 & 67
& 0.12 & 0.063 & 47 \\
Cut 1 & 0.22 & 0.13 & 1.89
& 0.052 & 0.033 & 0.36 \\
Cut 2 & 0.22 & 0.13 & 0.15~[0.29]
& 0.052 & 0.033 & 0.025~[0.061] \\
Cut 3 & 0.09 & 0.075 & 0.025~[0.052]
& 0.046 & 0.030 & 0.011~[0.030] \\
\bottomrule[1.5pt]
\end{tabular}
\end{table}

The cut-flow results are summarized in
Table~\ref{cutflow2}. After identifying
the associated-electron candidate through
the two-electron selection (Cut~1), the
mass-dependent energy requirement (Cut~2)
substantially suppresses the continuum SM
background while leaving the signal
essentially unchanged. The subsequent
fat-jet selection (Cut~3) provides further
rejection of the residual background. After the full selection, the background cross
sections are $0.025$~fb and $0.052$~fb for the $m_E=800$ and
$900$~GeV hypotheses at $\sqrt{s}=1$~TeV, and $0.011$~fb and
$0.030$~fb for $m_E=1300$ and $1400$~GeV at
$\sqrt{s}=1.5$~TeV, respectively.

\subsection{Case 3: $e^+e^-+b\bar b$}

For the Higgs channel, we focus on the dominant decay mode
$h\to b\bar b$.
The dominant SM background arises from
$e^+e^-\to e^+e^-jj$,
with contributions from $ZZ$, $Zh$, off-shell $Z/\gamma^*$ production,
and $t$-channel processes. Such events can contain two isolated
electrons together with a hadronic system that mimics the reconstructed
Higgs candidate.

\begin{figure}[htb]
\begin{center}
\centerline{\hspace{2.0cm}\epsfxsize=9cm\epsffile{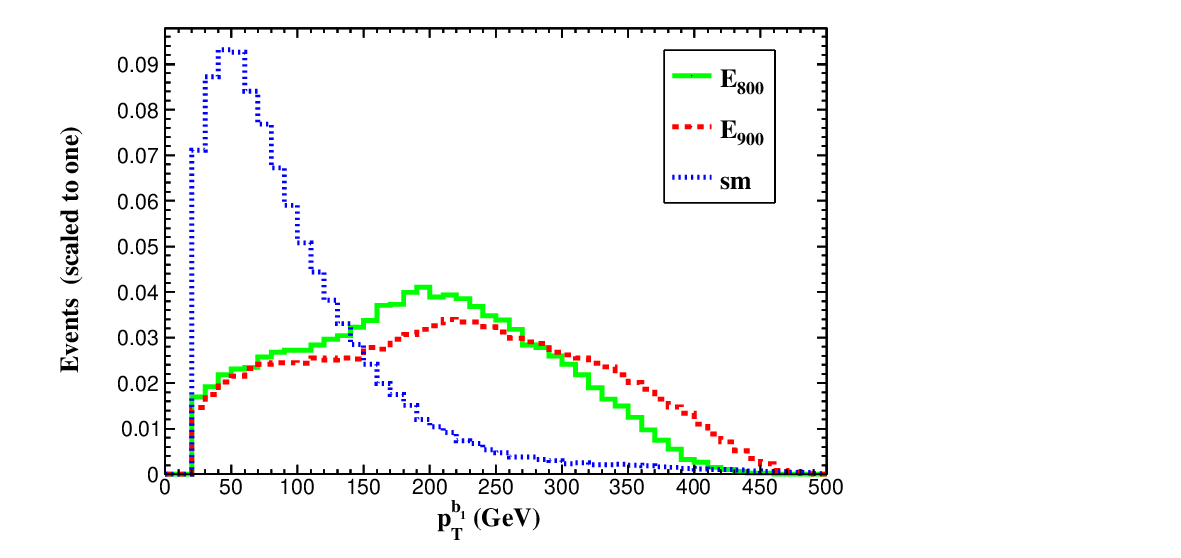}
\hspace{-2.0cm}\epsfxsize=9cm\epsffile{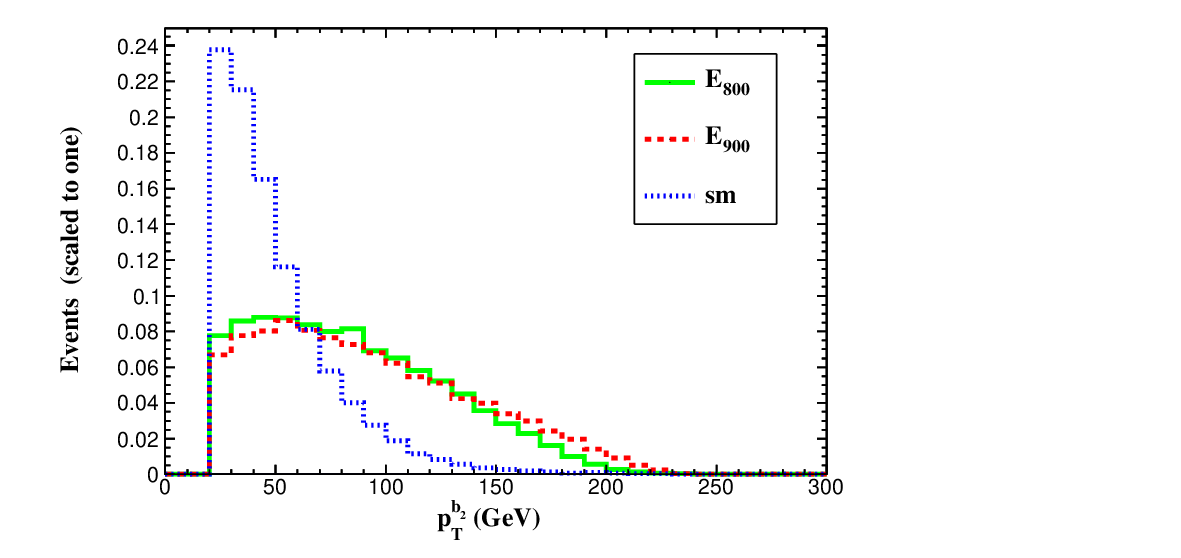}}
\centerline{\hspace{2.0cm}\epsfxsize=9cm\epsffile{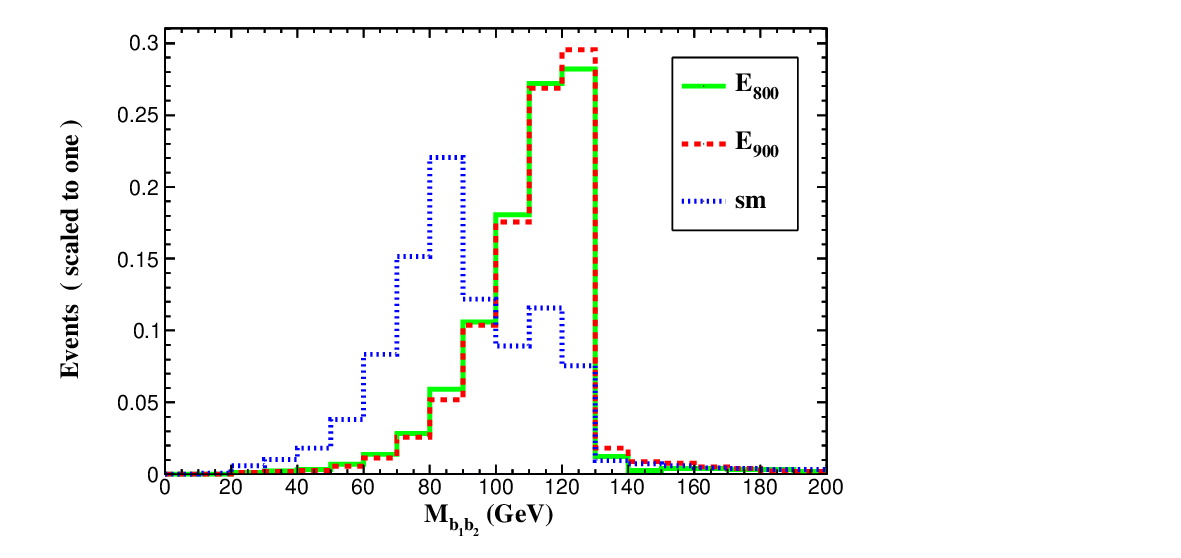}
\hspace{-2.0cm}\epsfxsize=9cm\epsffile{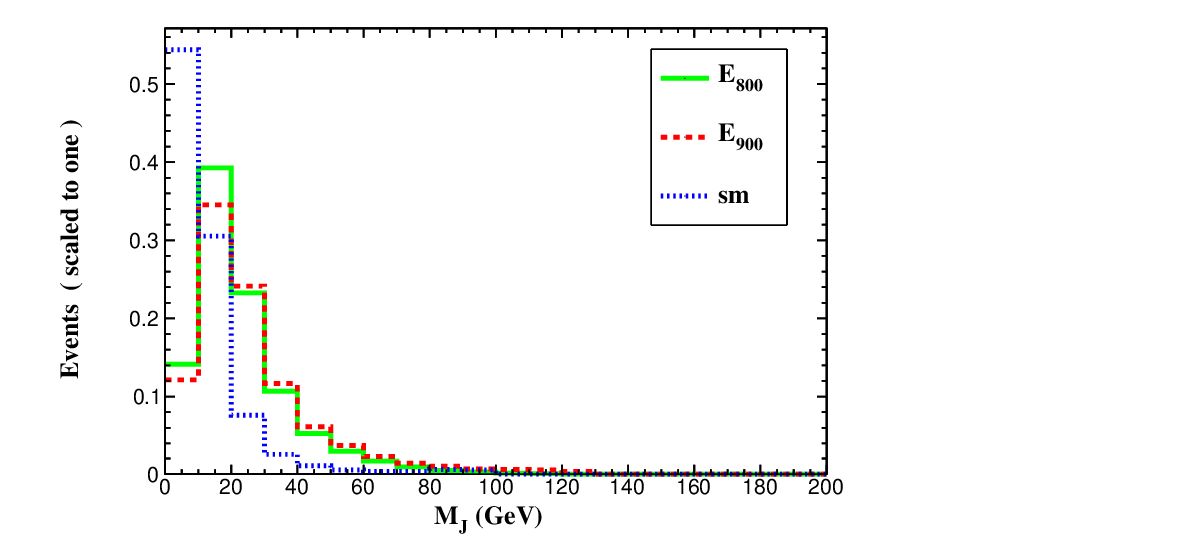}}
\caption{Normalized distributions of the main kinematic observables
for Case~3 and the dominant SM background at $\sqrt{s}=1$~TeV,
for $m_E=800$ and $900$~GeV. Shown are
$p_T^{b_1}$ and $p_T^{b_2}$ (top panels),
$M_{b_1b_2}$ (bottom left), and $M_J$ of the leading fat-jet candidate (bottom right).}
\label{fig7}
\end{center}
\end{figure}

\begin{figure}[ht]
\begin{center}
\centerline{\hspace{2.0cm}\epsfxsize=9cm\epsffile{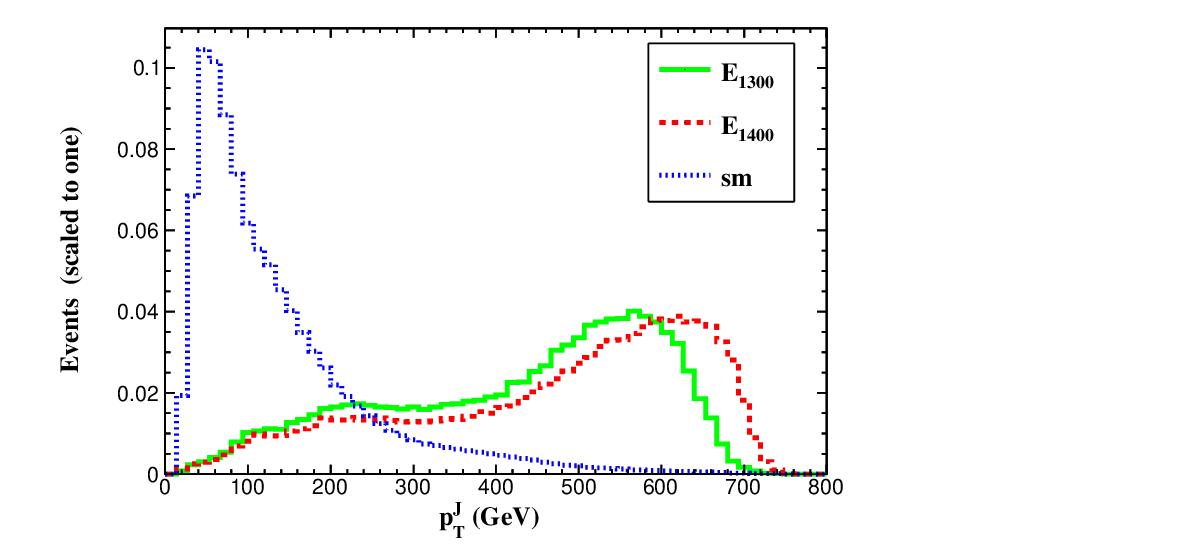}
\hspace{-2.0cm}\epsfxsize=9cm\epsffile{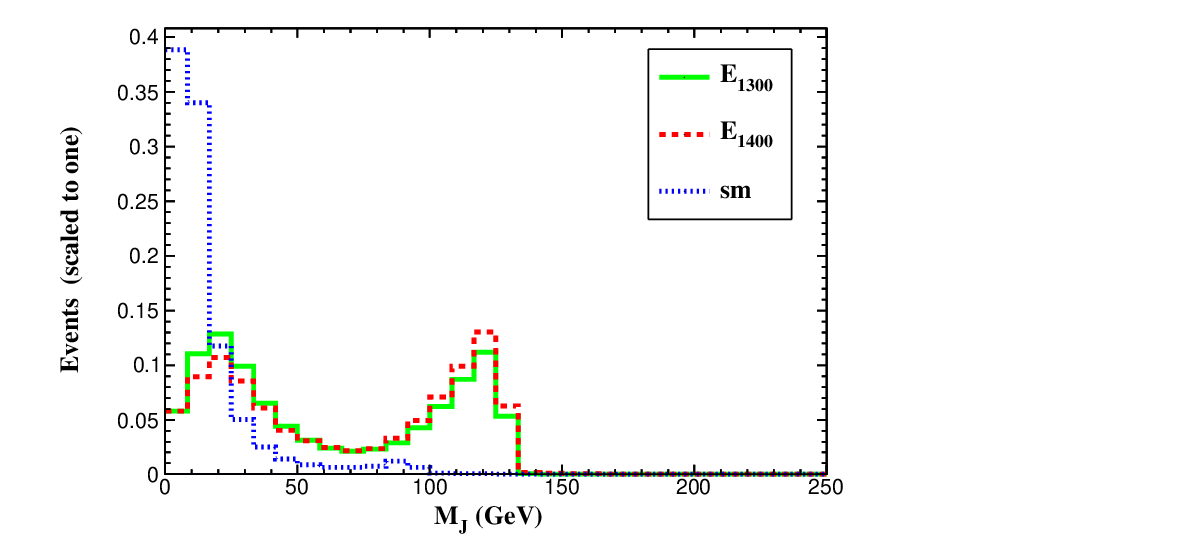}}
\caption{Normalized distributions of the transverse momentum $p_T^J$
(left) and invariant mass $M_J$ (right) of the leading fat-jet for
Case~3 at $\sqrt{s}=1.5$~TeV, with $m_E=1300$ and $1400$~GeV.}
\label{fig8}
\end{center}
\end{figure}

The appropriate Higgs reconstruction strategy depends on the collider
energy. At $\sqrt{s}=1$~TeV, for $m_E=800$ and $900$~GeV, the Higgs
boson is not sufficiently boosted for its two $b$-jets to merge into a
single fat-jet. The two $b$-jets therefore remain resolved, and, as
shown in Fig.~\ref{fig7}, their invariant mass $M_{b_1b_2}$ exhibits
a peak around the Higgs-boson mass. For comparison, Fig.~\ref{fig7} also shows the invariant mass $M_J$ of the
leading fat-jet candidate. Unlike the resolved $b\bar b$ invariant
mass, $M_J$ does not exhibit a comparably pronounced Higgs-mass peak
at $\sqrt{s}=1~{\rm TeV}$, further motivating the use of the resolved
$b$-jet reconstruction at this collider energy. At $\sqrt{s}=1.5$~TeV, for
$m_E=1300$ and $1400$~GeV, the Higgs boson is substantially more
boosted and its hadronic decay products are reconstructed as a single
fat-jet $J$. As shown in Fig.~\ref{fig8}, this fat-jet is characterized
by a large transverse momentum $p_T^J$ and an invariant mass $M_J$
peaked around $m_h=125$~GeV. At $\sqrt{s}=1.5$~TeV, the boosted Higgs
candidate is reconstructed using its fat-jet
transverse momentum and invariant mass,
without imposing a dedicated double-$b$-tagging
requirement. The corresponding sensitivity
should therefore be interpreted within this
mass-based boosted-Higgs reconstruction strategy.

The associated-electron energy $E_{e_{\rm assoc}}$ is not shown
separately in this channel. Its distribution is governed by the same
two-body production kinematics discussed in Sec.~\ref{sec:recoil} and
illustrated in Cases~1 and 2. We therefore apply the same
mass-hypothesis-dependent associated-electron energy selection in
Case~3.
The event selection follows the same electron-selection strategy as in
Case~2, while the Higgs reconstruction is adapted to the collider energy.

\begin{itemize}

\item \textbf{Cut 1:}
Exactly two isolated, oppositely charged electrons are required. The
leading electron must satisfy
$p_T^{e_1}>250$~GeV at $\sqrt{s}=1$~TeV and
$p_T^{e_1}>400$~GeV at $\sqrt{s}=1.5$~TeV, while the
sub-leading electron is required to have $p_T^{e_2}<100$~GeV.
In addition, the two electrons must be well separated,
with $\Delta R_{e_1,e_2}>1.0$.
The sub-leading electron $e_2$ is identified as the associated-electron
candidate.

\item \textbf{Cut 2:}
The associated-electron energy
$E_{e_2}$ is required to satisfy the mass-hypothesis-dependent
window specified in Eq.~(\ref{eq:associateenergywindow}), corresponding to
$|M_{\rm recoil}-m_E|<20$~GeV.

\item \textbf{Cut 3:}
The Higgs reconstruction strategy depends on the collider energy.
At $\sqrt{s}=1$~TeV, at least two resolved $b$-tagged jets
are required, with
$p_T^{b_1}>100$~GeV and $p_T^{b_2}>50$~GeV.
Their invariant mass must satisfy
$|M_{b_1b_2}-m_h|<25$~GeV.
At $\sqrt{s}=1.5$~TeV, the Higgs boson is reconstructed
using a boosted fat-jet $J$, which is required to satisfy
$p_T^J>200$~GeV and
$|M_J-m_h|<25$~GeV.

\end{itemize}

\begin{table}[htbp]
\centering
\small
\setlength{\tabcolsep}{4pt}
\caption{
Cut-flow cross sections (in fb) for the signal benchmarks and the SM
background in Case~3 at $\sqrt{s}=1$ and $1.5$~TeV, assuming
$\epsilon=0.1$. For Cuts~2 and~3, the unbracketed background entries
correspond to the $m_E=800$ and $1300$~GeV hypotheses at
$\sqrt{s}=1$ and $1.5$~TeV, respectively, while the entries in
square brackets correspond to the $m_E=900$ and $1400$~GeV
hypotheses.
}
\label{cutflow3}
\vspace{0.1cm}
\begin{tabular}{l ccc ccc}
\toprule[1.5pt]
 & \multicolumn{3}{c}{$\sqrt{s}=1$ TeV}
 & \multicolumn{3}{c}{$\sqrt{s}=1.5$ TeV} \\
\cmidrule{2-4} \cmidrule{5-7}
\multirow{2}{*}{Cuts}
 & \multicolumn{2}{c}{Signal} & Background
 & \multicolumn{2}{c}{Signal} & Background \\
 & 800~GeV & 900~GeV & $e^{+}e^{-}jj$
 & 1300~GeV & 1400~GeV & $e^{+}e^{-}jj$ \\
\midrule[0.8pt]
Basic & 0.68  & 0.31  & 67
      & 0.13  & 0.066 & 47 \\
Cut 1 & 0.29  & 0.18  & 1.89
      & 0.055 & 0.034 & 0.36 \\
Cut 2 & 0.29  & 0.18  & 0.146~[0.294]
      & 0.054 & 0.034 & 0.025~[0.061] \\
Cut 3 & 0.076 & 0.046 & 0.007~[0.0032]
      & 0.021 & 0.015 & 0.001~[0.0015] \\
\bottomrule[1.5pt]
\end{tabular}
\end{table}

The cut-flow results are summarized in Table~\ref{cutflow3}.
As in Case~2, the recoil-equivalent energy selection is imposed after
identifying the sub-leading electron and before the hadronic Higgs
selection. The associated-electron energy requirement in Cut~2 provides a further
substantial reduction of the SM background after the two-electron
selection, while the final Higgs reconstruction suppresses the remaining
background to the $10^{-3}$--$10^{-2}$~fb level. After all selections,
the background cross sections are $0.007$~fb and $0.0032$~fb for
$m_E=800$ and $900$~GeV at $\sqrt{s}=1$~TeV, and $0.001$~fb and
$0.0015$~fb for $m_E=1300$ and $1400$~GeV at
$\sqrt{s}=1.5$~TeV, respectively.

In Case~3, we focus exclusively on the dominant
$h\to b\bar b$ decay mode, which has a branching ratio
of approximately $58\%$.
Although the $h\to WW^*$ channel also has a sizable
branching fraction, it is not included as an additional
signal contribution in the present analysis.
At $\sqrt{s}=1$~TeV, the requirement of two resolved
$b$-tagged jets provides strong discrimination against
final states without a genuine $b\bar b$ pair.
At $\sqrt{s}=1.5$~TeV, the analysis instead employs
a boosted Higgs reconstruction strategy based on a
fat-jet, without considering a dedicated boosted
$h\to WW^*$ topology.
Therefore, the results presented for
Case~3 correspond specifically to the
$h\to b\bar b$ signal contribution
within the reconstruction and selection
strategy adopted in this study.

For all three channels, the mass-hypothesis-dependent associated-electron
energy selection is applied as early as the final-state topology permits.
This ordering exploits the two-body production kinematics before
reconstructing the boosted \(W\), \(Z\), or Higgs candidates, preserving the
signal efficiency while controlling the residual SM background.  The three
channels thus provide complementary probes of the first-generation singlet
VLL scenario within the same baseline cut-based framework.
\subsection{Statistical Analysis}
We estimate the expected discovery and exclusion
sensitivities using the Asimov median-significance
approach~\cite{Cowan:2010js}:
\begin{equation}
Z_{\mathrm{disc}}
=
\sqrt{
2\left[
(s+b)\ln\left(1+\frac{s}{b}\right)-s
\right]
},
\end{equation}
and
\begin{equation}
Z_{\mathrm{excl}}
=
\sqrt{
2\left[
s-b\ln\left(1+\frac{s}{b}\right)
\right]
}.
\end{equation}
Here, $s$ and $b$ denote the expected
numbers of signal and background events,
respectively, for a given integrated
luminosity $\mathcal{L}$. The projected sensitivities are evaluated
including statistical uncertainties only.
A quantitative assessment of experimental
and theoretical systematic uncertainties
is beyond the scope of the present study.

The signal samples used in the cut-flow analysis are generated at the
reference coupling
$\epsilon_0=0.1$.
For each fixed VLL mass benchmark, the signal cross section at a
general value of the coupling $\epsilon$ is obtained using the
approximate scaling
\begin{equation}
\sigma_S(\epsilon)
=
\sigma_S(\epsilon_0)
\left(
\frac{\epsilon}{\epsilon_0}
\right)^2 .
\end{equation}
The corresponding value of the left-handed mixing angle is then
obtained from
\begin{equation}
\sin\theta_L
\simeq
\frac{\epsilon v}{m_E}.
\end{equation}
Thus, for each fixed VLL mass benchmark, the sensitivity obtained in
terms of $\epsilon$ is mapped onto the
$(\mathcal{L},\sin\theta_L)$ plane using the above relation.
The selection efficiencies and branching fractions are taken to be
independent of the overall value of $\epsilon$ within the prompt-decay
regime considered here. The expected signal yield is therefore given by
$s=\mathcal{L}\,\sigma_S(\epsilon)$.

\begin{figure}[b]
\begin{center}
\centerline{\hspace{1.0cm}\epsfxsize=7cm\epsffile{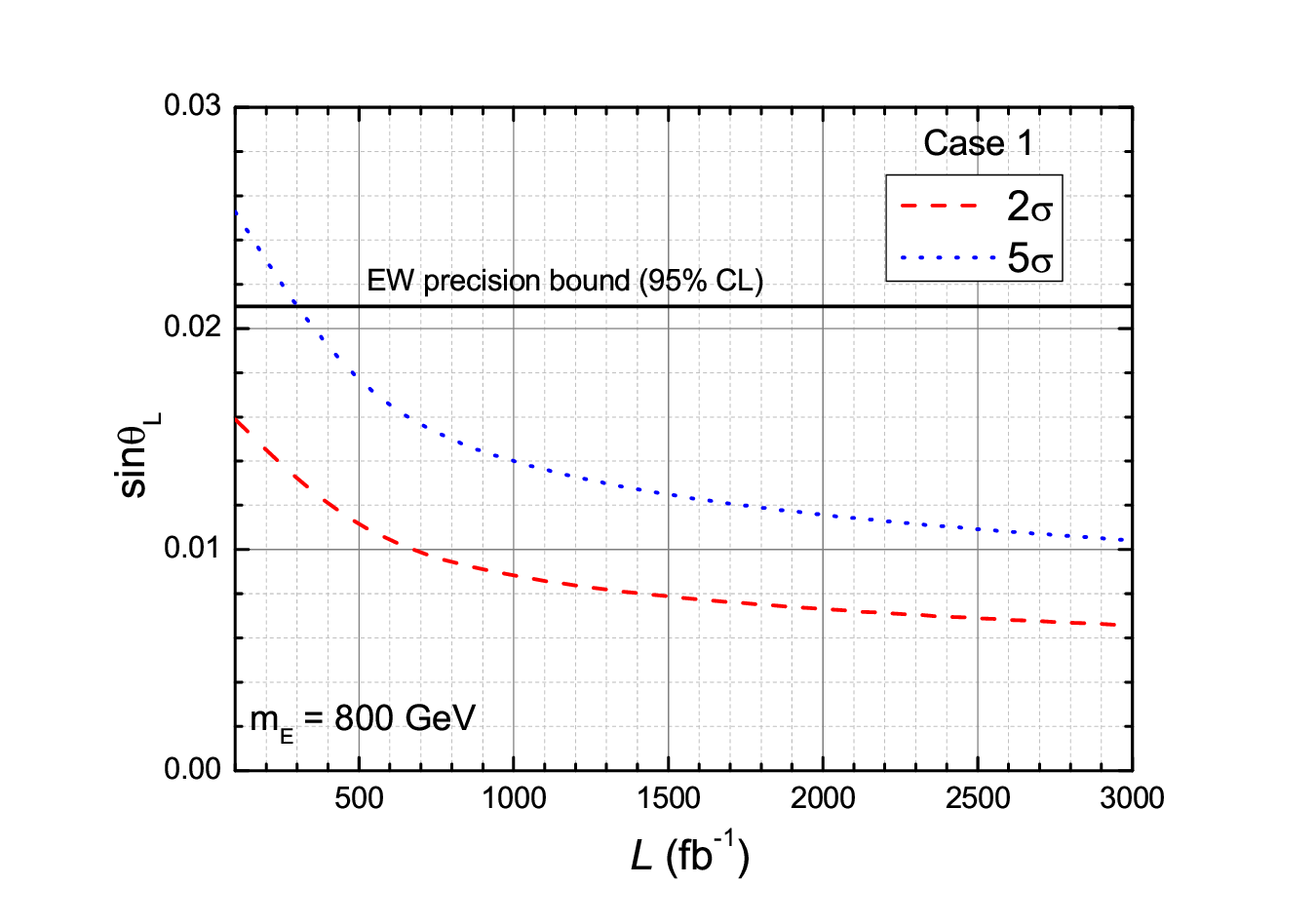}
\hspace{-1.0cm}\epsfxsize=7cm\epsffile{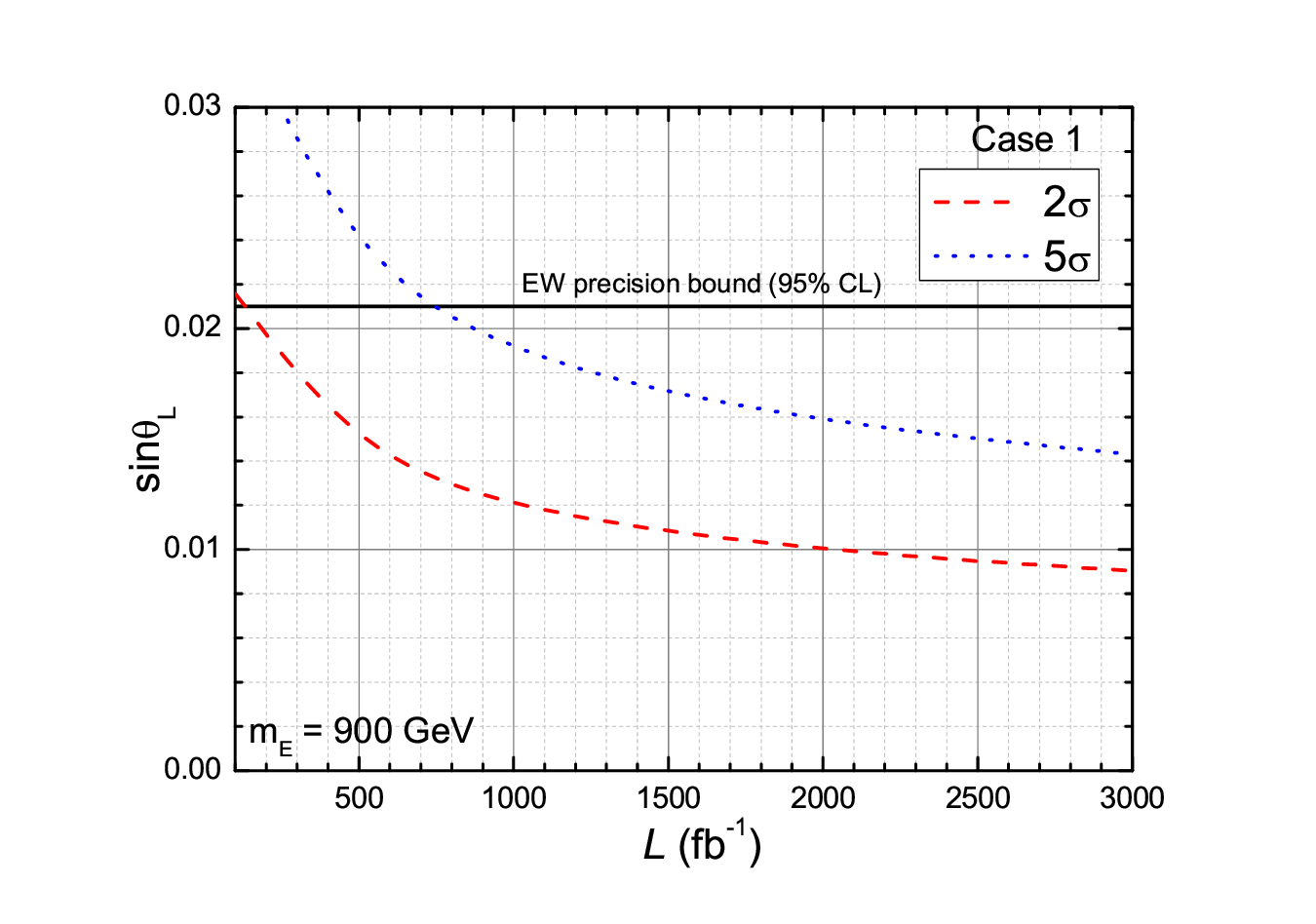}}
\centerline{\hspace{1.0cm}\epsfxsize=7cm\epsffile{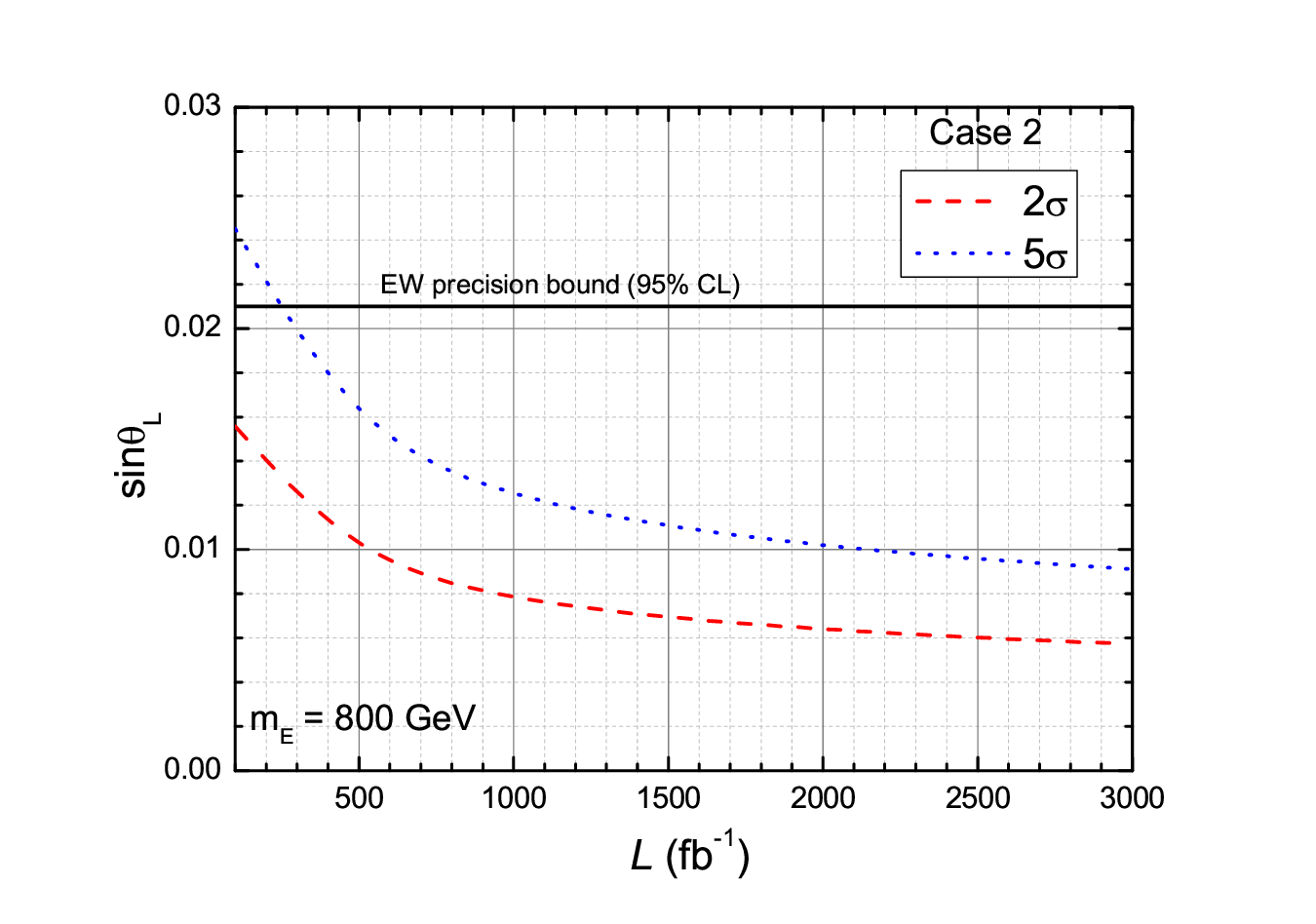}
\hspace{-1.0cm}\epsfxsize=7cm\epsffile{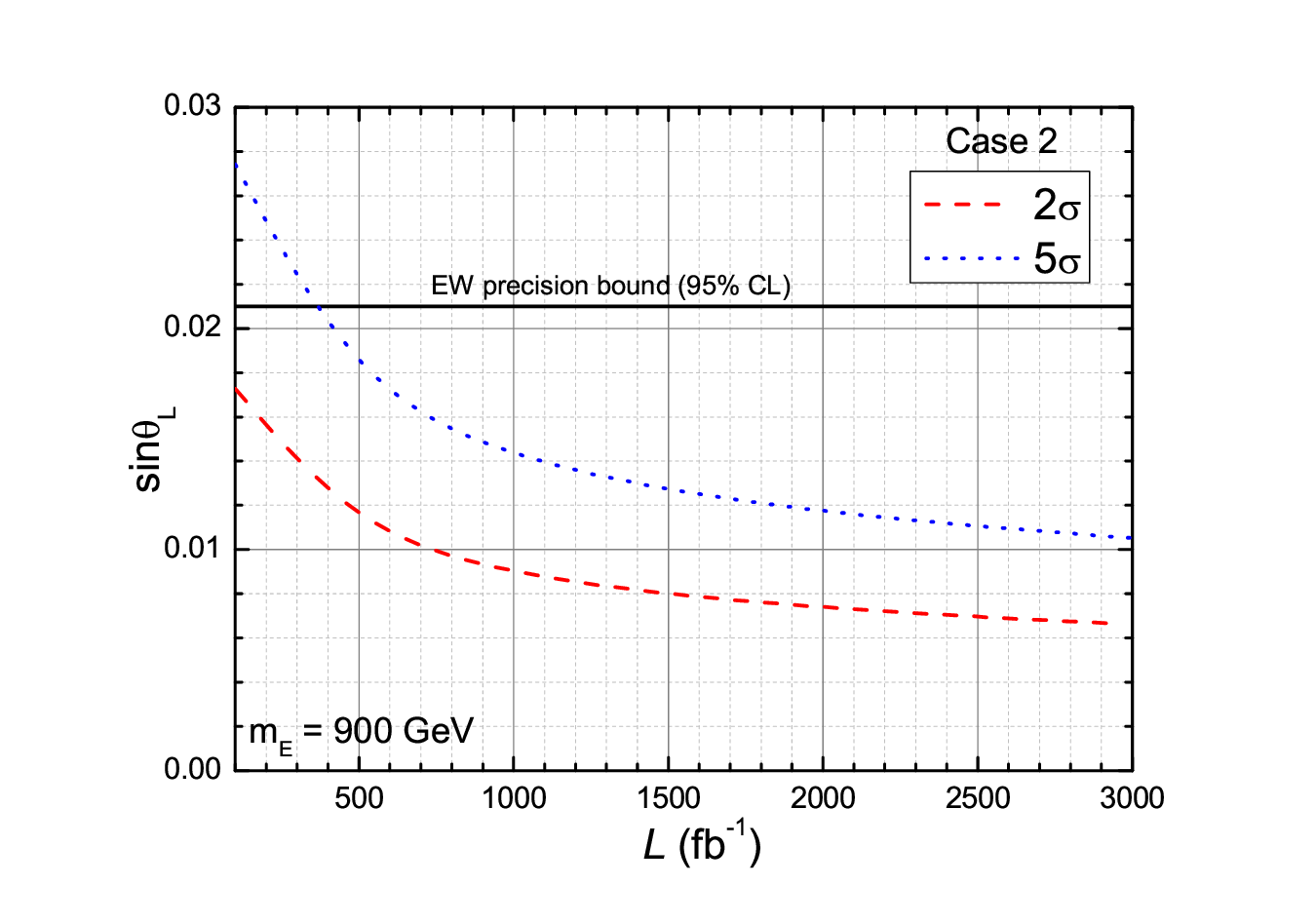}}
\centerline{\hspace{1.0cm}\epsfxsize=7cm\epsffile{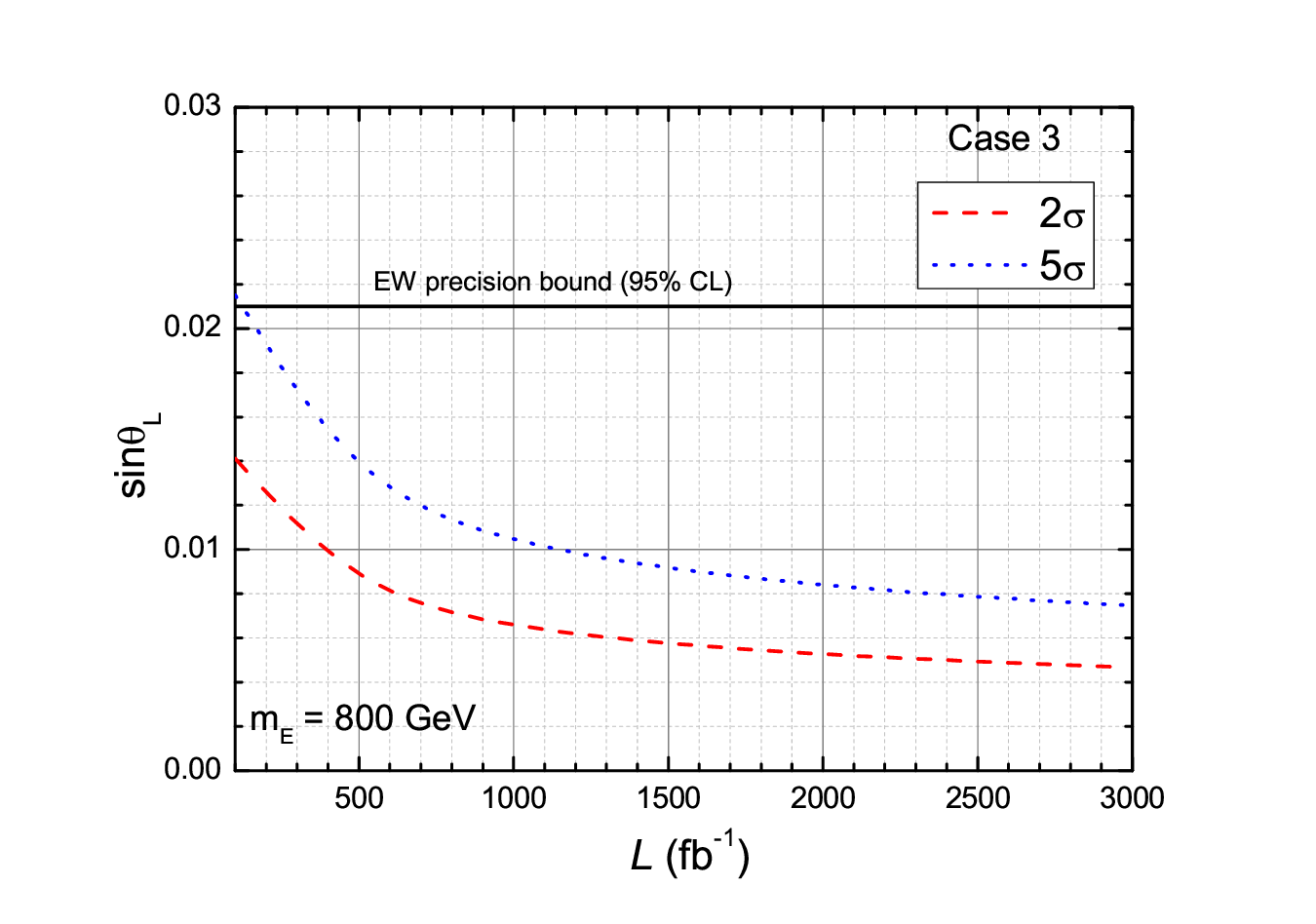}
\hspace{-1.0cm}\epsfxsize=7cm\epsffile{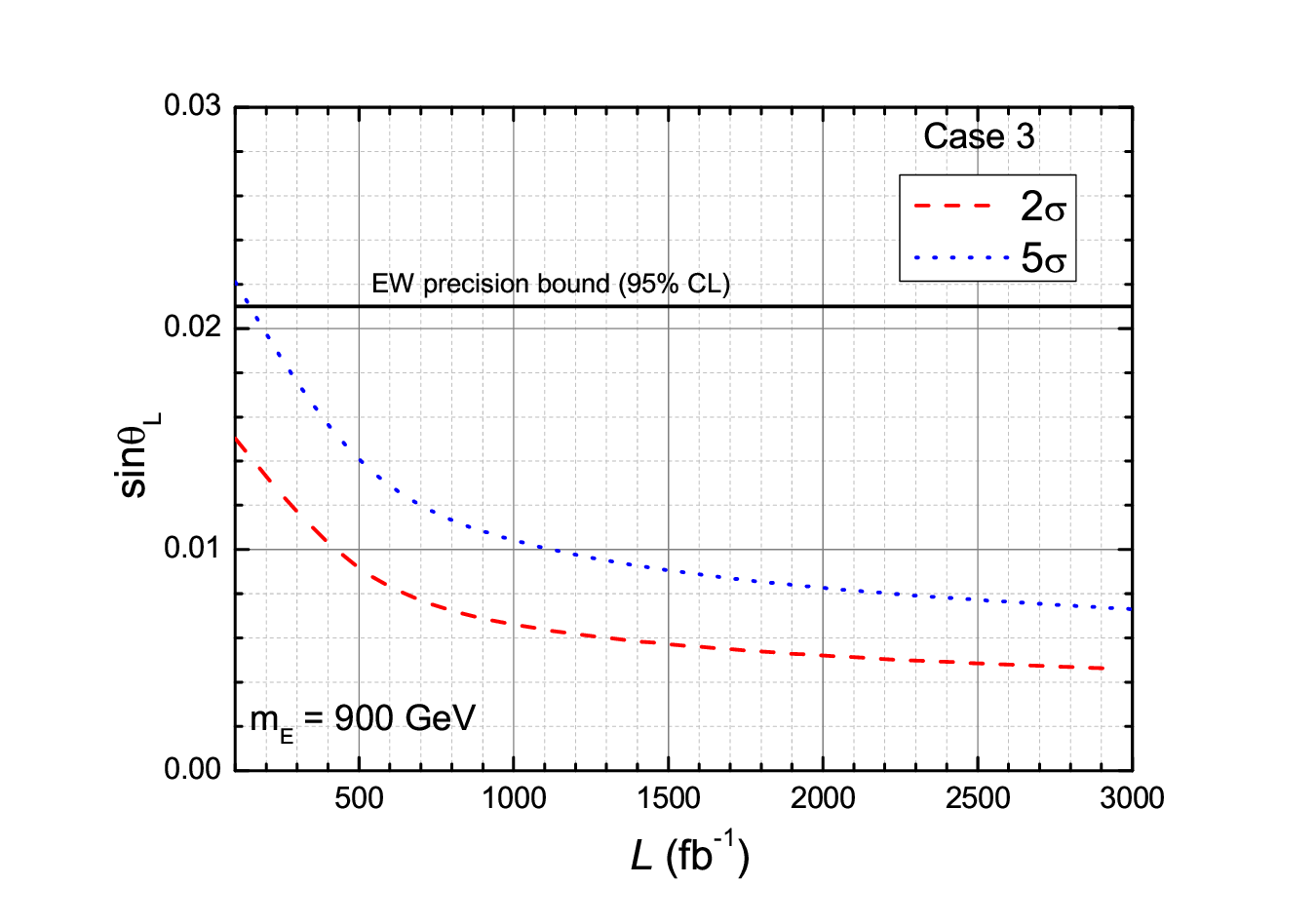}}
\caption{$2\sigma$ exclusion (dashed) and $5\sigma$ discovery
reaches (dotted) in the $(\mathcal{L},\sin\theta_L)$ plane at
$\sqrt{s}=1$~TeV for Case~1 (top), Case~2 (middle), and Case~3
(bottom). The left and right panels correspond to $m_E=800$ and
$900$~GeV, respectively. The horizontal solid line denotes the reference 95\% CL electroweak
precision constraint adopted from Ref.~\cite{deBlas:2013gla}.}
\label{fig9}
\end{center}
\end{figure}

\begin{figure}[!t]
\begin{center}
\centerline{\hspace{2.0cm}\epsfxsize=7.5cm\epsffile{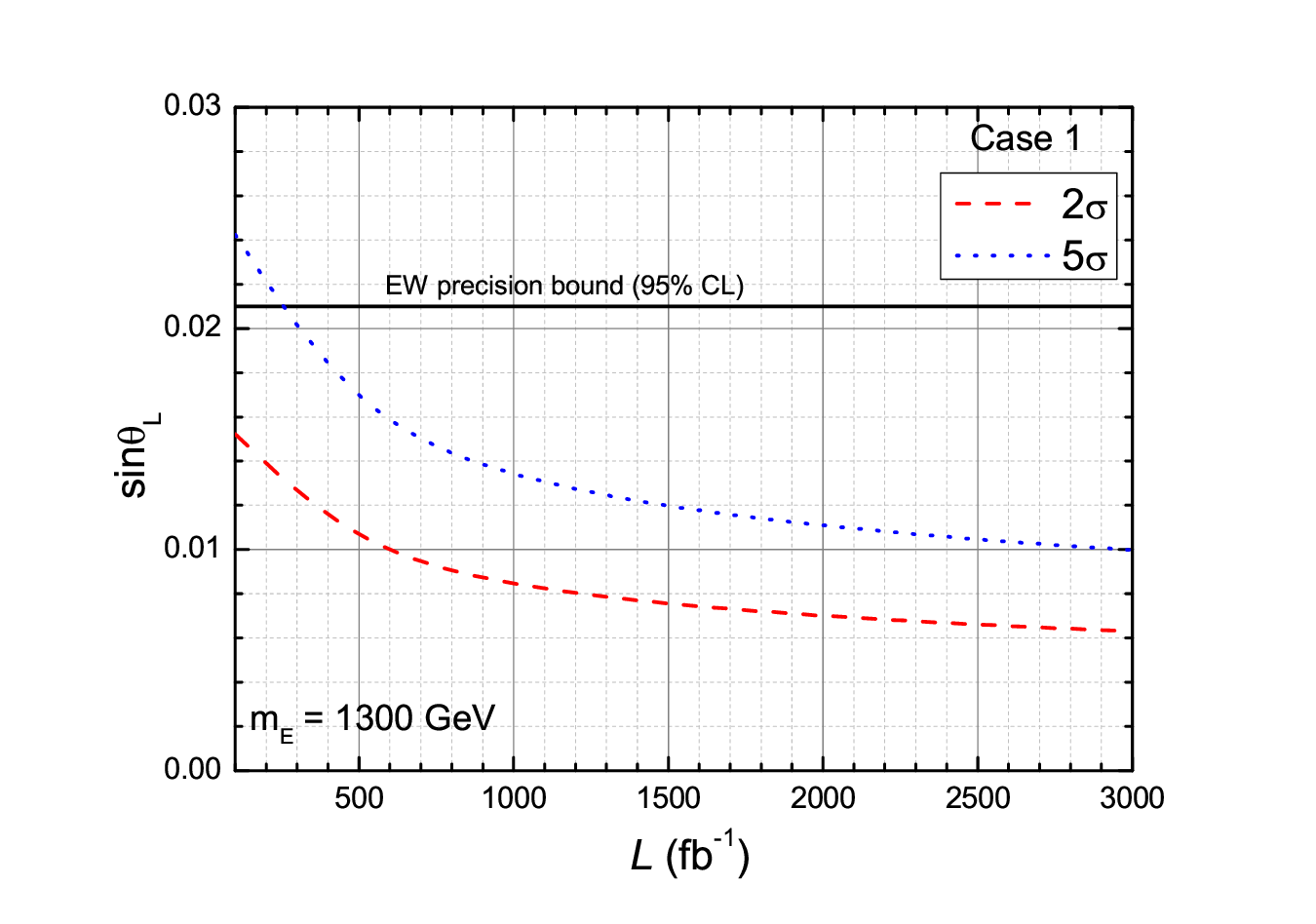}
\hspace{-1.0cm}\epsfxsize=7.5cm\epsffile{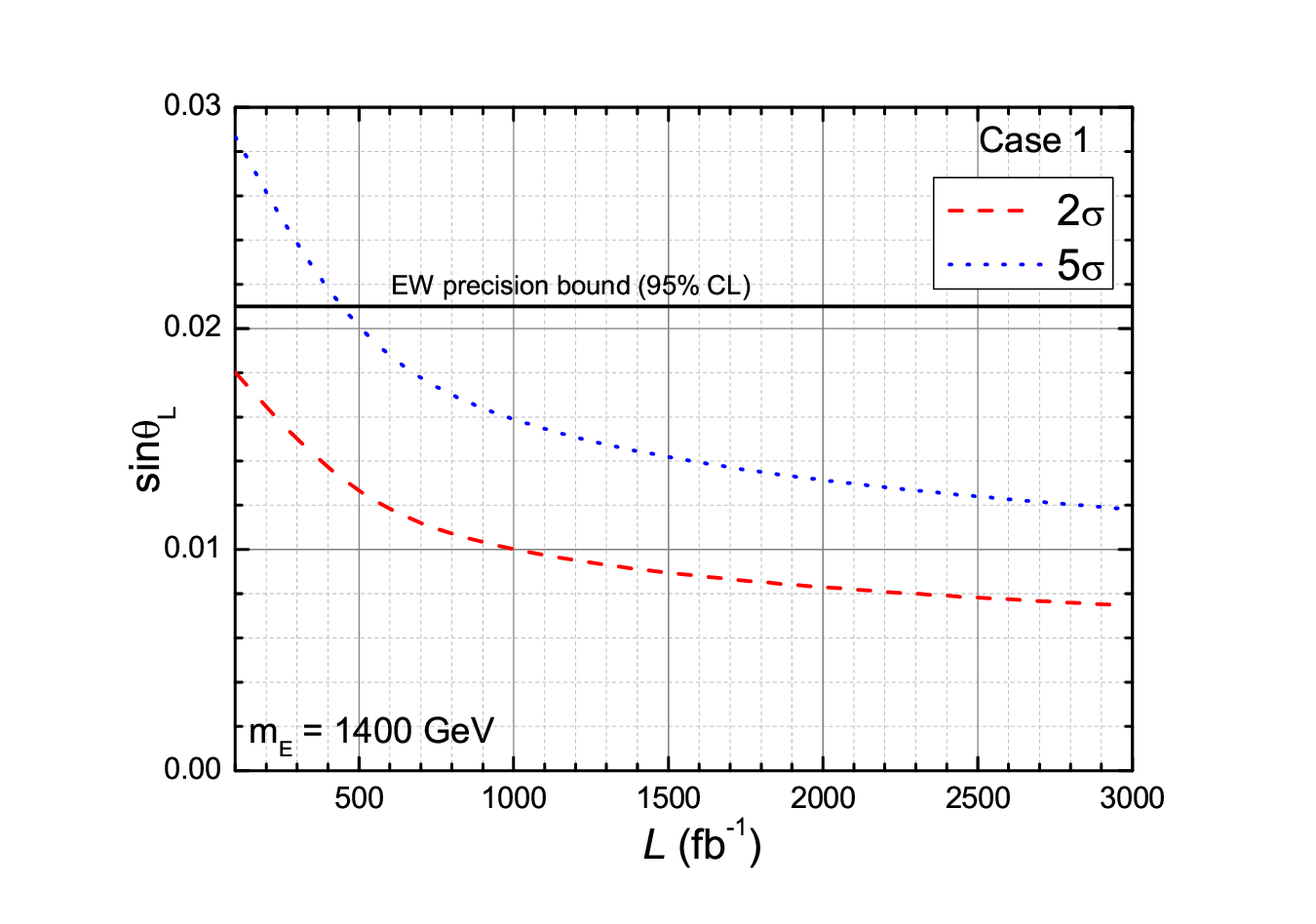}}
\centerline{\hspace{2.0cm}\epsfxsize=7.5cm\epsffile{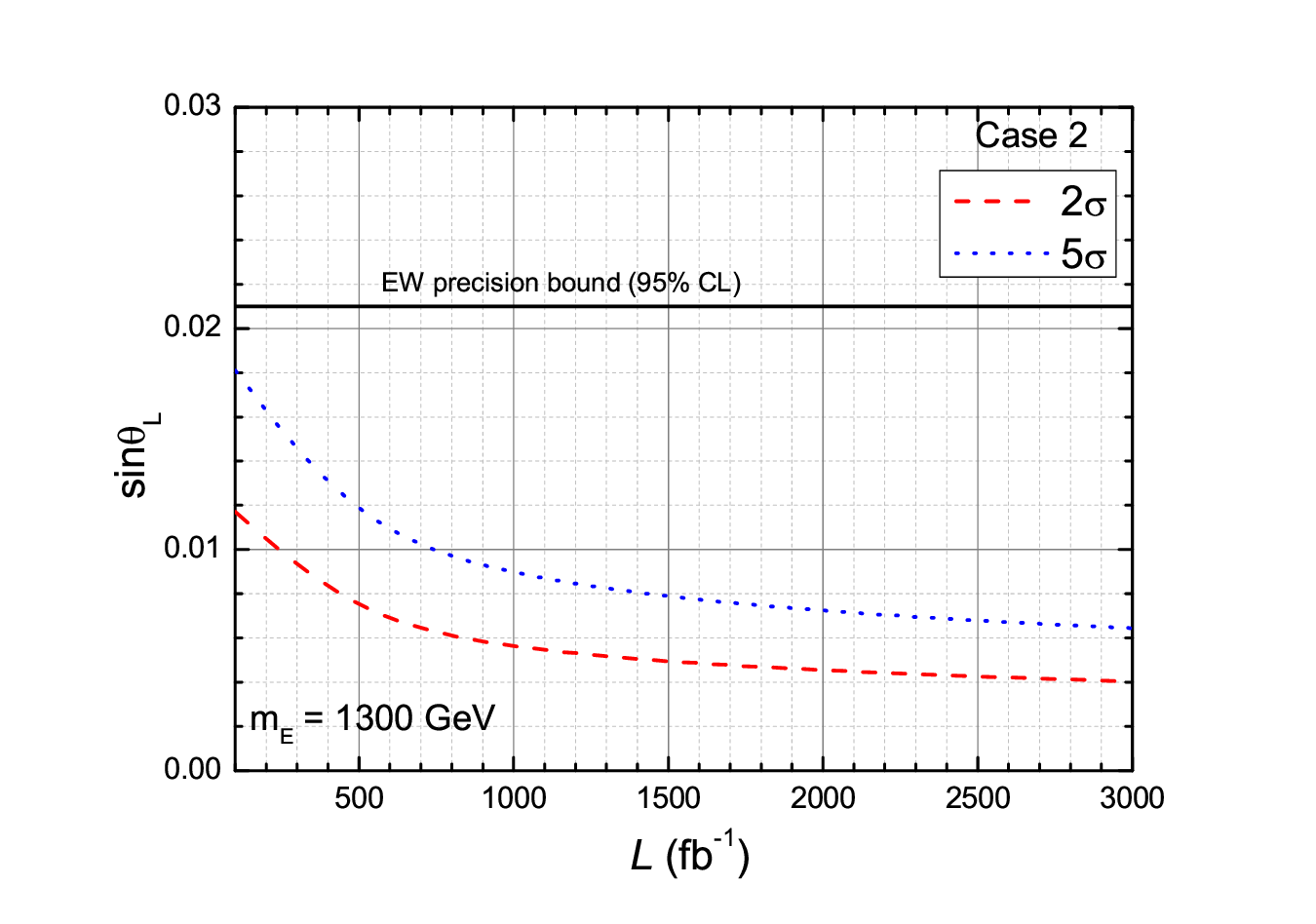}
\hspace{-1.0cm}\epsfxsize=7.5cm\epsffile{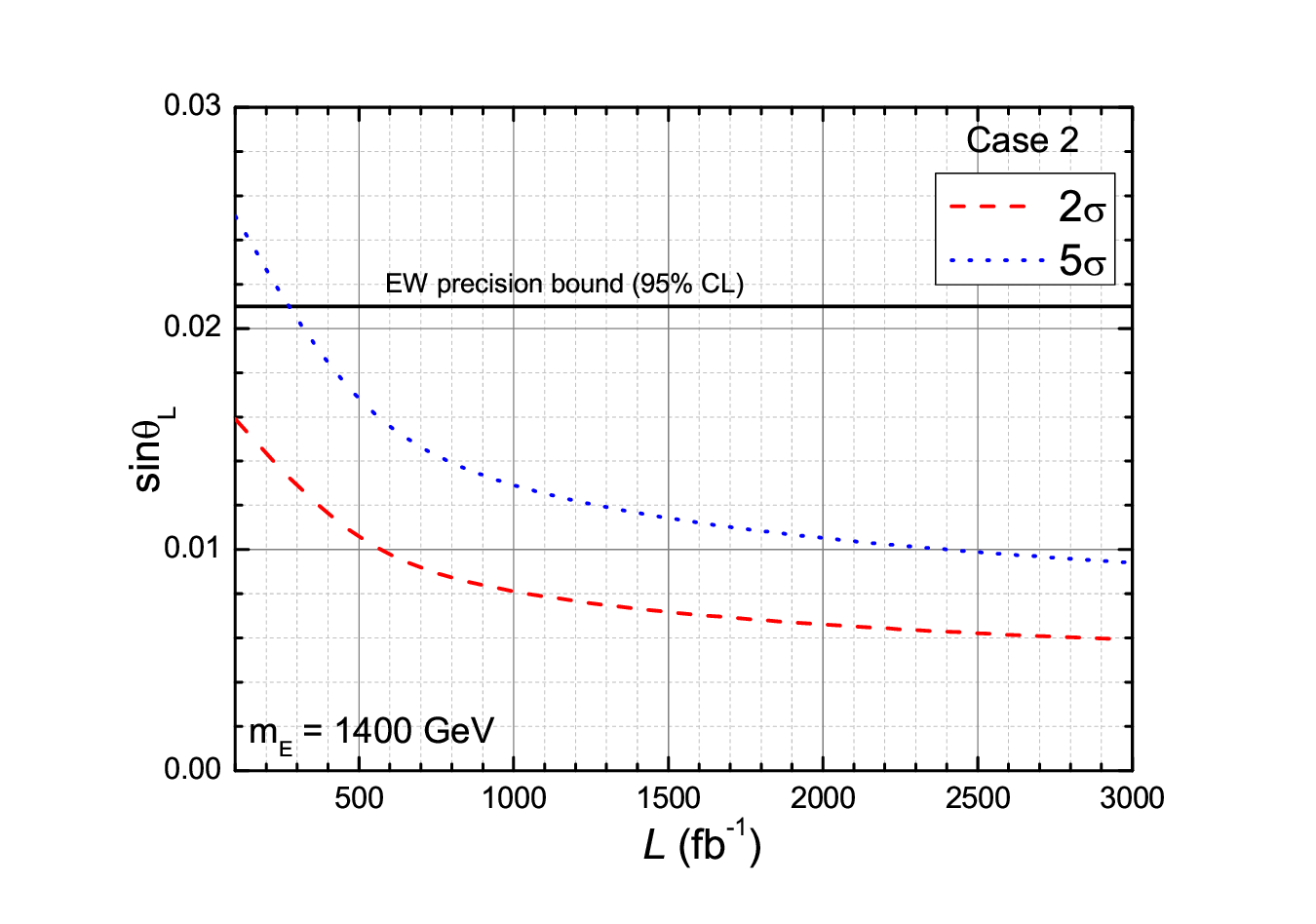}}
\centerline{\hspace{2.0cm}\epsfxsize=7.5cm\epsffile{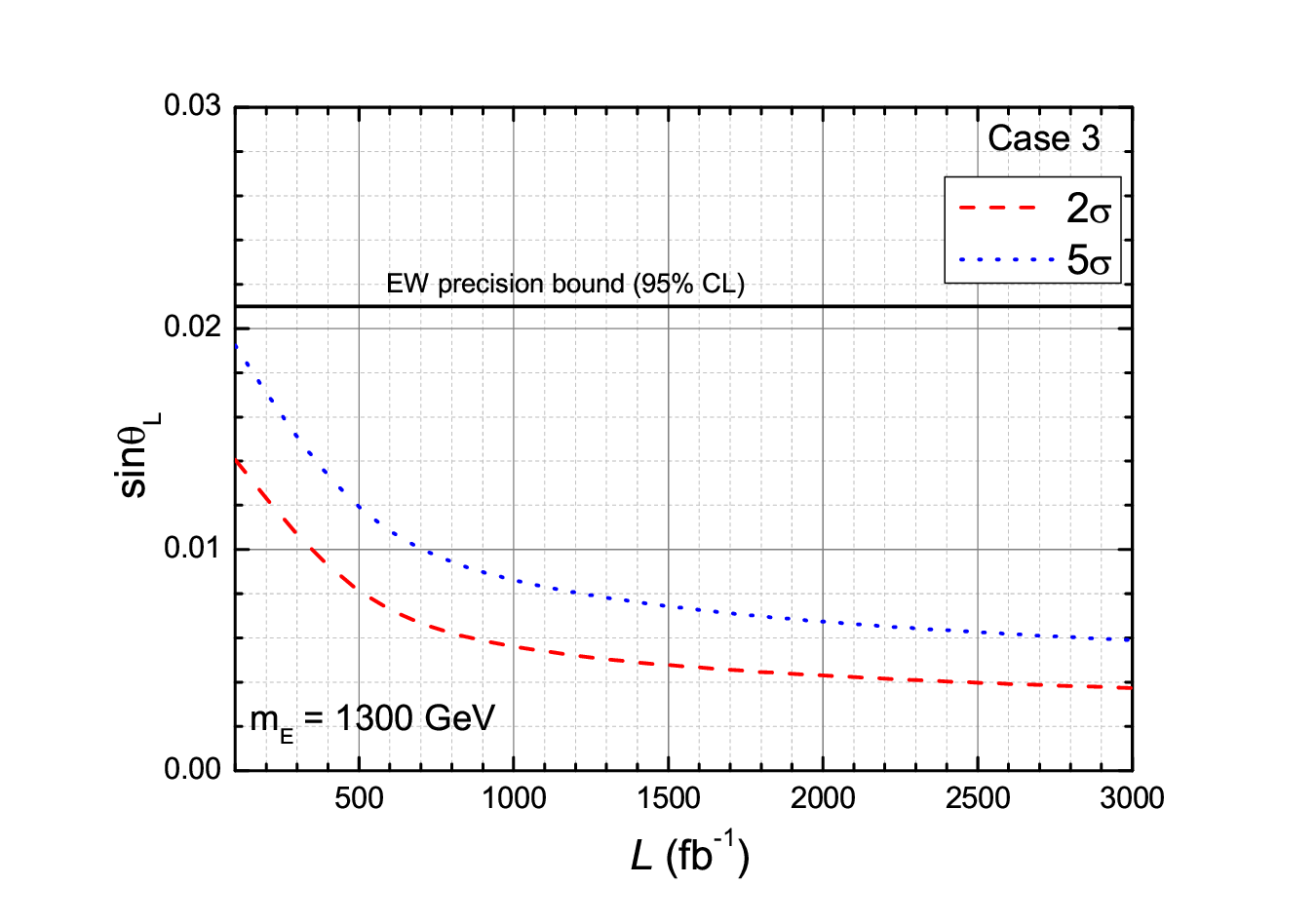}
\hspace{-1.0cm}\epsfxsize=7.5cm\epsffile{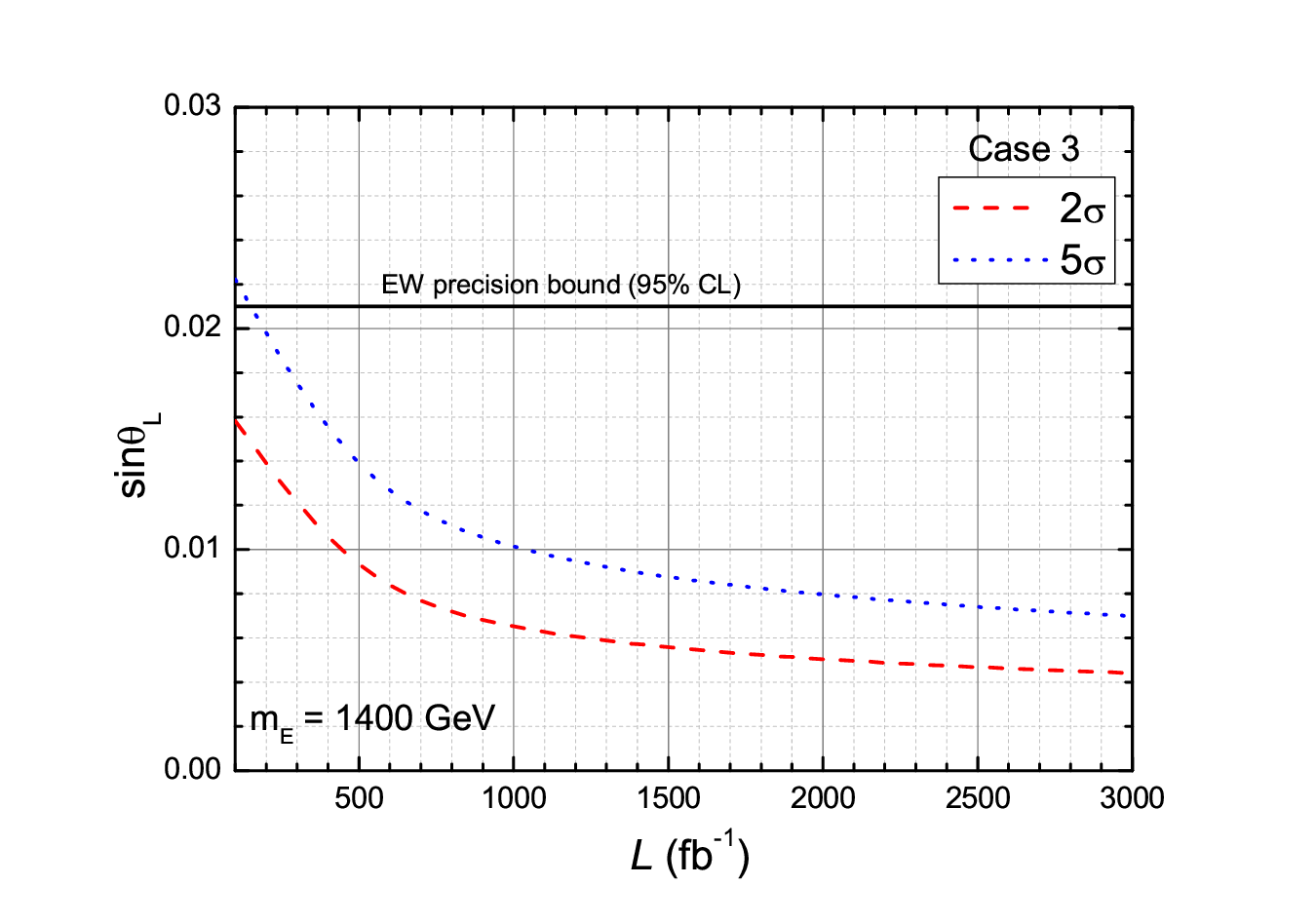}}
\caption{Same as Fig.~\ref{fig9}, but for $\sqrt{s}=1.5$~TeV with
$m_E=1300$~GeV (left) and $1400$~GeV (right).}
\label{fig10}
\end{center}
\end{figure}

The $2\sigma$ exclusion and $5\sigma$ discovery reaches on
$\sin\theta_L$ are shown in Figs.~\ref{fig9} and~\ref{fig10} as
functions of the integrated luminosity $\mathcal{L}$ at
$\sqrt{s}=1$ and $1.5$~TeV, respectively. In each panel, the dashed
and dotted curves denote the $2\sigma$ exclusion and $5\sigma$
discovery reaches, while the horizontal solid line indicates the reference 95\% CL
electroweak precision constraint,
$\sin\theta_L\lesssim0.021$, adopted from Ref.~\cite{deBlas:2013gla}.

For the $2\sigma$ exclusion reach, relatively modest integrated
luminosities of a few hundred fb$^{-1}$ are sufficient, for the
benchmark masses considered here, to probe values of $\sin\theta_L$
well below the reference electroweak precision bound adopted above.
This demonstrates that direct searches at future $e^+e^-$ colliders
can extend the sensitivity beyond the reference indirect constraint
used for comparison in the present study.

For the $5\sigma$ discovery reach, the integrated luminosity required
to access a given value of $\sin\theta_L$, or equivalently the
accessible value of $\sin\theta_L$ at fixed luminosity, depends on
both the VLL mass and the decay channel. At
$\sqrt{s}=1$~TeV, the Higgs channel (Case~3) provides the strongest
projected sensitivity for both $m_E=800$ and $900$~GeV. This
performance results from the efficient reconstruction of the Higgs
candidate together with the strong suppression of the residual SM
background. The $eZ$ channel (Case~2) gives the next-best sensitivity,
while the $W\nu$ channel (Case~1) is comparatively less sensitive
within the present cut-based analysis.

At $\sqrt{s}=1.5$~TeV, Case~3 likewise provides the strongest
projected sensitivity for both benchmark masses, followed by Case~2
and Case~1. The relative separation among the three channels depends
on $m_E$ and reflects the combined effects of the signal efficiency,
the reconstruction of the visible decay products, and the residual SM
backgrounds after the full event selection. This ordering is specific to the present cut-based analysis strategy
and should not be interpreted as an intrinsic hierarchy among the
three VLL decay modes; further channel-dependent optimization could
modify their relative sensitivities.

Overall, the three decay channels provide complementary probes of the
first-generation singlet VLL scenario.  For all benchmark masses considered, the projected sensitivity extends
to values of $\sin\theta_L$ below the reference electroweak precision
constraint adopted from Ref.~\cite{deBlas:2013gla}.
 Future improvements in electroweak precision measurements may lead to
stronger indirect constraints on the VLL mixing parameter. Direct
collider searches nevertheless provide complementary information by
probing on-shell VLL production together with its characteristic decay
signatures. The two approaches therefore test the VLL parameter space
in qualitatively different ways and can provide complementary
sensitivity at future $e^+e^-$ facilities.

\section{Conclusions and Discussion}
\label{sec:conclusion}

In this work, we have performed a detailed Monte Carlo study of the
single production of first-generation weak-isosinglet VLLs at future
$e^+e^-$ colliders with $\sqrt{s}=1~{\rm TeV}$ and
$1.5~{\rm TeV}$ within a baseline cut-based analysis framework.
We consider three signal channels,
$e^+e^-\to e^\pm E^\mp$ with
$E^\pm\to W^\pm\nu_e$ (Case~1),
$E^\pm\to e^\pm Z$ (Case~2),
and $E^\pm\to e^\pm h$ (Case~3).
For Cases~1 and 2, as well as Case~3 at
$\sqrt{s}=1.5~{\rm TeV}$, the boosted hadronic decays of the
$W$, $Z$, and Higgs bosons are reconstructed as single fat-jets.
For Case~3 at $\sqrt{s}=1~{\rm TeV}$, the Higgs boson is instead
reconstructed from two resolved $b$-tagged jets.
A beam-polarization configuration of $-80\%$ for electrons and
$+30\%$ for positrons is adopted to enhance the signal production
rate. We evaluate the $2\sigma$ exclusion and $5\sigma$ discovery reaches
on $\sin\theta_L$ in the $(\mathcal{L},\sin\theta_L)$ plane, using
the 95\% CL bound
$\sin\theta_L\lesssim0.021$ from Ref.~\cite{deBlas:2013gla} as a reference
electroweak precision constraint.

Our main findings can be summarized as follows.

\begin{itemize}

\item
For the $2\sigma$ exclusion reach, relatively modest integrated
luminosities of a few hundred ${\rm fb}^{-1}$ are sufficient, for the
benchmark masses considered here, to probe values of $\sin\theta_L$ well below the reference
electroweak precision bound adopted in this study. This illustrates the potential of direct searches at future lepton
colliders to access parameter regions beyond the reference indirect
constraint used for comparison.

\item
The $5\sigma$ discovery reach depends on both the VLL mass and the
decay channel. At $\sqrt{s}=1~{\rm TeV}$, the Higgs channel
(Case~3) provides the strongest projected sensitivity for the
benchmark masses $m_E=800$ and $900~{\rm GeV}$, followed by the
$eZ$ channel (Case~2) and the $W\nu$ channel (Case~1).
At $\sqrt{s}=1.5~{\rm TeV}$, Case~3 likewise provides the strongest
projected sensitivity for both $m_E=1300$ and $1400~{\rm GeV}$,
followed by Cases~2 and 1. These relative differences reflect the
combined effects of the signal efficiency, the reconstruction of the
visible decay products, and the residual SM backgrounds after the full
event selection.

\item
The mass-hypothesis-dependent associated-electron energy selection,
formulated through the recoil-mass relation, provides a common and
decay-channel-independent strategy for the three signal topologies
considered here. Combined with the reconstruction of the $W$, $Z$,
and Higgs decay products, this approach provides complementary
sensitivity across the three channels. A statistical combination of
these channels could further improve the overall sensitivity.

\end{itemize}

It should be emphasized that the sensitivities presented in this work
are obtained within a baseline cut-based analysis strategy. Systematic
uncertainties are not included, and ISR, beamstrahlung, and realistic
luminosity-spectrum effects are not incorporated in either the
production rates or the event-level kinematics. The results should
therefore be interpreted as baseline sensitivity estimates at fixed
nominal center-of-mass energy. The mass-hypothesis-dependent associated-electron energy selection,
motivated by the recoil-mass relation, already exploits the
characteristic two-body kinematics at the nominal center-of-mass
energy. Further improvements may
nevertheless be achievable through more sophisticated kinematic
reconstruction, multivariate techniques, channel-dependent
optimization, or a combined analysis of the three decay channels.
Further selection optimization should be assessed using the
final statistical sensitivity rather than signal acceptance alone.

For the benchmark configurations investigated here, sizeable
sensitivity is retained for VLL masses as high as
$m_E=900~{\rm GeV}$ at $\sqrt{s}=1~{\rm TeV}$ and
$m_E=1400~{\rm GeV}$ at $\sqrt{s}=1.5~{\rm TeV}$.
These benchmark points lie close to the corresponding
single-production kinematic boundaries and demonstrate the potential
of future $e^+e^-$ colliders to probe heavy first-generation singlet
VLLs. A quantitative determination of the ultimate mass reach would,
however, require a dedicated scan in $m_E$ together with a more fully
optimized analysis. The projected sensitivities also extend to values
of $\sin\theta_L$ below the reference electroweak precision bound over
substantial regions of the parameter space, highlighting the
complementarity between direct collider searches and indirect
precision constraints.

Although we do not perform a dedicated HL-LHC projection, the present
results illustrate the potential of future lepton colliders to probe
small VLL mixing angles through clean single-production signatures.
Higher-energy hadron colliders, such as the FCC-hh, could provide
complementary information through pair-production processes, while
higher-energy stages of future lepton colliders, including a possible
$\sqrt{s}=3~{\rm TeV}$ CLIC stage, could further extend the accessible
VLL mass range. Dedicated studies of these scenarios would be valuable
for establishing the broader discovery potential for first-generation
singlet VLLs.
\begin{acknowledgments}
\vspace*{-0.5truecm}
Y.-B. L.  is supported by the Natural Science Foundation of Henan Province (Grant No. 252300421988).
S. M. is supported  through the NExT Institute and STFC CG ST/X000583/1.
\end{acknowledgments}


\begin{thebibliography}{99}
\bibitem{Graham:2009gy}
P.~W.~Graham, A.~Ismail, S.~Rajendran and P.~Saraswat,
\href{doi:10.1103/PhysRevD.81.055016}{Phys.\ Rev.\ D \textbf{81}, 055016 (2010)}.
[arXiv:0910.3020 [hep-ph]].

\bibitem{Endo:2011xq}
M.~Endo, K.~Hamaguchi, S.~Iwamoto and N.~Yokozaki,
\href{doi:10.1103/PhysRevD.85.095012}{Phys.\ Rev.\ D \textbf{85}, 095012 (2012)}.
[arXiv:1112.5653 [hep-ph]].

\bibitem{Martin:2012dg}
S.~P.~Martin and J.~D.~Wells,
\href{doi:10.1103/PhysRevD.86.035017}{Phys.\ Rev.\ D \textbf{86}, 035017 (2012)}.
[arXiv:1206.2956 [hep-ph]].

\bibitem{Endo:2012cc}
M.~Endo, K.~Hamaguchi, K.~Ishikawa, S.~Iwamoto and N.~Yokozaki,
\href{doi:10.1007/JHEP01(2013)181}{JHEP \textbf{01}, 181 (2013)}.
[arXiv:1212.3935 [hep-ph]].

\bibitem{Fischler:2013tva}
W.~Fischler and W.~Tangarife,
\href{doi:10.1007/JHEP05(2014)151}{JHEP \textbf{05}, 151 (2014)}.
[arXiv:1310.6369 [hep-ph]].
\bibitem{Huang:2012kz}
G.~Y.~Huang, K.~Kong and S.~C.~Park,
\href{doi:10.1007/JHEP06(2012)099}{JHEP \textbf{06}, 099 (2012)}.
[arXiv:1204.0522 [hep-ph]].

\bibitem{Kong:2010qd}
K.~Kong, S.~C.~Park and T.~G.~Rizzo,
\href{doi:10.1007/JHEP07(2010)059}{JHEP \textbf{07}, 059 (2010)}.
[arXiv:1004.4635 [hep-ph]].



\bibitem{DeCurtis:2018iqd}
S.~De Curtis, L.~Delle Rose, S.~Moretti and K.~Yagyu,
\href{doi:10.1016/j.physletb.2018.09.042}{Phys. Lett. B \textbf{786}, 189 (2018)}.
[arXiv:1803.01865 [hep-ph]].

\bibitem{He:1999vp}
H.~J.~He, T.~M.~P.~Tait and C.~P.~Yuan,
\href{doi:10.1103/PhysRevD.62.011702}{Phys.\ Rev.\ D \textbf{62}, 011702(R) (2000)}.
[arXiv:hep-ph/9911266 [hep-ph]].

\bibitem{Wang:2013jwa}
X.~F.~Wang, C.~Du and H.~J.~He,
\href{doi:10.1016/j.physletb.2013.05.015}{Phys.\ Lett.\ B \textbf{723}, 314-323 (2013)}.
[arXiv:1304.2257 [hep-ph]].

\bibitem{He:2001fz}
H.~J.~He, C.~T.~Hill and T.~M.~P.~Tait,
\href{doi:10.1103/PhysRevD.65.055006}{Phys.\ Rev.\ D \textbf{65}, 055006 (2002)}.
[arXiv:hep-ph/0108041 [hep-ph]].

\bibitem{He:2014ora}
H.~J.~He and Z.~Z.~Xianyu,
\href{doi:10.1088/1475-7516/2014/10/019}{JCAP \textbf{10}, 019 (2014)}.
[arXiv:1405.7331 [hep-ph]].

\bibitem{Arkani-Hamed:2012dcq}
N.~Arkani-Hamed, K.~Blum, R.~T.~D'Agnolo and J.~Fan,
\href{doi:10.1007/JHEP01(2013)149}{JHEP \textbf{01}, 149 (2013)}.
[arXiv:1207.4482 [hep-ph]].

\bibitem{Xiao:2014kba}
M.~L.~Xiao and J.~H.~Yu,
\href{doi:10.1103/PhysRevD.90.014007}{Phys.\ Rev.\ D \textbf{90}, no.1, 014007 (2014)}.
[arXiv:1404.0681 [hep-ph]].

\bibitem{Cingiloglu:2024vdh}
K.~Y.~Cingiloglu and M.~Frank,
\href{doi:10.1103/PhysRevD.111.016025}{Phys.\ Rev.\ D \textbf{111}, no.1, 016025 (2025)}.
[arXiv:2408.10898 [hep-ph]].
\bibitem{Schwaller:2013hqa}
P.~Schwaller, T.~M.~P.~Tait and R.~Vega-Morales,
\href{doi:10.1103/PhysRevD.88.035001}{Phys.\ Rev.\ D \textbf{88}, no.3, 035001 (2013)}.
[arXiv:1305.1108 [hep-ph]].

\bibitem{Halverson:2014nwa}
J.~Halverson, N.~Orlofsky and A.~Pierce,
\href{doi:10.1103/PhysRevD.90.015002}{Phys.\ Rev.\ D \textbf{90}, no.1, 015002 (2014)}.
[arXiv:1403.1592 [hep-ph]].

\bibitem{Bahrami:2016has}
S.~Bahrami, M.~Frank, D.~K.~Ghosh, N.~Ghosh and I.~Saha,
\href{doi:10.1103/PhysRevD.95.095024}{Phys.\ Rev.\ D \textbf{95}, no.9, 095024 (2017)}.
[arXiv:1612.06334 [hep-ph]].

\bibitem{Bhattacharya:2018fus}
S.~Bhattacharya, P.~Ghosh, N.~Sahoo and N.~Sahu,
\href{doi:10.3389/fphy.2019.00080}{Front.\ in Phys.\ \textbf{7}, 80 (2019)}.
[arXiv:1812.06505 [hep-ph]].



\bibitem{Hiller:2019mou}
G.~Hiller, C.~Hormigos-Feliu, D.~F.~Litim and T.~Steudtner,
\href{doi:10.1103/PhysRevD.102.071901}{Phys.\ Rev.\ D \textbf{102}, no.7, 071901 (2020)}.
[arXiv:1910.14062 [hep-ph]].

\bibitem{DeJesus:2020yqx}
A.~S.~De Jesus, S.~Kovalenko, F.~S.~Queiroz, C.~Siqueira and K.~Sinha,
\href{doi:10.1103/PhysRevD.102.035004}{Phys.\ Rev.\ D \textbf{102}, no.3, 035004 (2020)}.
[arXiv:2004.01200 [hep-ph]].

\bibitem{Frank:2020smf}
M.~Frank and I.~Saha,
\href{doi:10.1103/PhysRevD.102.115034}{Phys.\ Rev.\ D \textbf{102}, no.11, 115034 (2020)}.
[arXiv:2008.11909 [hep-ph]].

\bibitem{Dermisek:2021ajd}
R.~Dermisek, K.~Hermanek and N.~McGinnis,
\href{doi:10.1103/PhysRevD.104.055033}{Phys.\ Rev.\ D \textbf{104}, no.5, 055033 (2021)}.
[arXiv:2103.05645 [hep-ph]].

\bibitem{Brune:2022rlo}
T.~Brune, T.~W.~Kephart and H.~P\"as,
\href{doi:10.1140/epjc/s10052-024-13617-5}{Eur.\ Phys.\ J.\ C \textbf{84}, no.12, 1254 (2024)}.
[arXiv:2205.05566 [hep-ph]].

\bibitem{Guedes:2022cfy}
G.~Guedes and P.~Olgoso,
\href{doi:10.1007/JHEP09(2022)181}{JHEP \textbf{09}, 181 (2022)}.
[arXiv:2205.04480 [hep-ph]].

\bibitem{Erdelyi:2025axy}
B.~A.~Erdelyi, R.~Gr{\"o}ber and N.~Selimovic,
\href{doi:10.1007/JHEP05(2025)135}{JHEP \textbf{05}, 135 (2025)}.
[arXiv:2501.07628 [hep-ph]].
\bibitem{Athron:2025ets}
P.~Athron, K.~M{\"o}hling, D.~St{\"o}ckinger and H.~St{\"o}ckinger-Kim,
\href{doi:10.1016/j.ppnp.2025.104225}{Prog. Part. Nucl. Phys. \textbf{148}, 104225 (2026)}.
[arXiv:2507.09289 [hep-ph]].

\bibitem{He:2022zjz}
S.~P.~He,
\href{doi:10.1088/1674-1137/ac9e4c}{Chin.\ Phys.\ C \textbf{47}, no.4, 043102 (2023)}.
[arXiv:2205.02088 [hep-ph]].

\bibitem{Kawamura:2022fhm}
J.~Kawamura and S.~Raby,
\href{doi:10.1103/PhysRevD.106.035009}{Phys.\ Rev.\ D \textbf{106}, no.3, 035009 (2022)}.
[arXiv:2205.10480 [hep-ph]].

\bibitem{delAguila:2008pw}
F.~del Aguila, J.~de Blas and M.~Perez-Victoria,
\href{doi:10.1103/PhysRevD.78.013010}{Phys.\ Rev.\ D \textbf{78}, 013010 (2008)}.
[arXiv:0803.4008 [hep-ph]].

\bibitem{Ishiwata:2013gma}
K.~Ishiwata and M.~B.~Wise,
\href{doi:10.1103/PhysRevD.88.055009}{Phys.\ Rev.\ D \textbf{88}, 055009 (2013)}.
[arXiv:1307.1112 [hep-ph]].

\bibitem{Kearney:2012zi}
J.~Kearney, A.~Pierce and N.~Weiner,
\href{doi:10.1103/PhysRevD.86.113005}{Phys.\ Rev.\ D \textbf{86}, 113005 (2012)}.
[arXiv:1207.7062 [hep-ph]].

\bibitem{Altmannshofer:2013zba}
W.~Altmannshofer, M.~Bauer and M.~Carena,
\href{doi:10.1007/JHEP01(2014)060}{JHEP \textbf{01}, 060 (2014)}.
[arXiv:1308.1987 [hep-ph]].

\bibitem{Falkowski:2013jya}
A.~Falkowski, D.~M.~Straub and A.~Vicente,
\href{doi:10.1007/JHEP05(2014)092}{JHEP \textbf{05}, 092 (2014)}.
[arXiv:1312.5329 [hep-ph]].

\bibitem{Ellis:2014dza}
S.~A.~R.~Ellis, R.~M.~Godbole, S.~Gopalakrishna and J.~D.~Wells,
\href{doi:10.1007/JHEP09(2014)130}{JHEP \textbf{09}, 130 (2014)}.
[arXiv:1404.4398 [hep-ph]].

\bibitem{Ishiwata:2015cga}
K.~Ishiwata, Z.~Ligeti and M.~B.~Wise,
\href{doi:10.1007/JHEP10(2015)027}{JHEP \textbf{10}, 027 (2015)}.
[arXiv:1506.03484 [hep-ph]].

\bibitem{Dermisek:2014cia}
R.~Dermisek, A.~Raval and S.~Shin,
\href{doi:10.1103/PhysRevD.90.034023}{Phys.\ Rev.\ D \textbf{90}, 034023 (2014)}.
[arXiv:1406.7018 [hep-ph]].

\bibitem{Crivellin:2020ebi}
A.~Crivellin, F.~Kirk, C.~A.~Manzari and M.~Montull,
\href{doi:10.1007/JHEP12(2020)166}{JHEP \textbf{12}, 166 (2020)}.
[arXiv:2008.01113 [hep-ph]].

\bibitem{Endo:2020tkb}
M.~Endo and S.~Mishima,
\href{doi:10.1007/JHEP08(2020)004}{JHEP \textbf{08}, 004 (2020)}.
[arXiv:2005.03933 [hep-ph]].

\bibitem{Chakrabarty:2020jro}
N.~Chakrabarty,
\href{doi:10.1140/epjp/s13360-021-02168-3}{Eur.\ Phys.\ J.\ Plus \textbf{136}, 1183 (2021)}.
[arXiv:2010.05215 [hep-ph]].

\bibitem{Guedes:2021oqx}
G.~Guedes and J.~Santiago,
\href{doi:10.1007/JHEP01(2022)111}{JHEP \textbf{01}, 111 (2022)}.
[arXiv:2107.03429 [hep-ph]].

\bibitem{Raju:2022zlv}
M.~Raju, A.~Mukherjee and J.~P.~Saha,
\href{doi:10.1140/epjc/s10052-023-11595-8}{Eur.\ Phys.\ J.\ C \textbf{83}, 429 (2023)}.
[arXiv:2207.02825 [hep-ph]].

\bibitem{Li:2023mrw}
X.~Q.~Li, Z.~J.~Xie, Y.~D.~Yang and X.~B.~Yuan,
\href{doi:10.1016/j.nuclphysb.2024.116646}{Nucl.\ Phys.\ B \textbf{1006}, 116646 (2024)}.
[arXiv:2307.05290 [hep-ph]].

\bibitem{Dermisek:2014qca}
R.~Dermisek, J.~P.~Hall, E.~Lunghi and S.~Shin,
\href{doi:10.1007/JHEP12(2014)013}{JHEP \textbf{12}, 013 (2014)}.
[arXiv:1408.3123 [hep-ph]].

\bibitem{Dermisek:2015oja}
R.~Dermisek, E.~Lunghi and S.~Shin,
\href{doi:10.1007/JHEP02(2016)119}{JHEP \textbf{02}, 119 (2016)}.
[arXiv:1509.04292 [hep-ph]].

\bibitem{Kumar:2015tna}
N.~Kumar and S.~P.~Martin,
\href{doi:10.1103/PhysRevD.92.115018}{Phys.\ Rev.\ D \textbf{92}, 115018 (2015)}.
[arXiv:1510.03456 [hep-ph]].

\bibitem{Dermisek:2015hue}
R.~Dermisek, E.~Lunghi and S.~Shin,
\href{doi:10.1007/JHEP05(2016)148}{JHEP \textbf{05}, 148 (2016)}.
[arXiv:1512.07837 [hep-ph]].

\bibitem{Chen:2016lsr}
C.~H.~Chen and T.~Nomura,
\href{doi:10.1140/epjc/s10052-016-4197-3}{Eur.\ Phys.\ J.\ C \textbf{76}, 353 (2016)}.
[arXiv:1602.07519 [hep-ph]].

\bibitem{Kawamura:2019rth}
J.~Kawamura, S.~Raby and A.~Trautner,
\href{doi:10.1103/PhysRevD.100.055030}{Phys.\ Rev.\ D \textbf{100}, 055030 (2019)}.
[arXiv:1906.11297 [hep-ph]].

\bibitem{Freitas:2020ttd}
F.~F.~Freitas, J.~Gon\c{c}alves, A.~P.~Morais and R.~Pasechnik,
\href{doi:10.1007/JHEP01(2021)076}{JHEP \textbf{01}, 076 (2021)}.
[arXiv:2010.01307 [hep-ph]].

\bibitem{OsmanAcar:2021plv}
A.~Osman Acar, O.~E.~Delialioglu and S.~Sultansoy,
[arXiv:2103.08222 [hep-ph]].

\bibitem{Kawamura:2021ygg}
J.~Kawamura and S.~Raby,
\href{doi:10.1103/PhysRevD.104.035007}{Phys.\ Rev.\ D \textbf{104}, 035007 (2021)}.
[arXiv:2104.04461 [hep-ph]].

\bibitem{Bonilla:2021ize}
C.~Bonilla, A.~E.~C\'arcamo Hern\'andez, J.~Gon\c{c}alves, F.~F.~Freitas, A.~P.~Morais and R.~Pasechnik,
\href{doi:10.1007/JHEP01(2022)154}{JHEP \textbf{01}, 154 (2022)}.
[arXiv:2107.14165 [hep-ph]].

\bibitem{Baspehlivan:2022qet}
F.~Baspehlivan, B.~Dagli, O.~E.~Delialioglu and S.~Sultansoy,
[arXiv:2201.08251 [hep-ph]].

\bibitem{Cao:2023smj}
Q.~H.~Cao, J.~Guo, J.~Liu, Y.~Luo and X.~P.~Wang,
\href{doi:10.1103/PhysRevD.110.015029}{Phys.\ Rev.\ D \textbf{110}, 015029 (2024)}.
[arXiv:2311.12934 [hep-ph]].

\bibitem{Bernreuther:2023uxh}
E.~Bernreuther and B.~A.~Dobrescu,
\href{doi:10.1007/JHEP07(2023)079}{JHEP \textbf{07}, 079 (2023)}.
[arXiv:2304.08509 [hep-ph]].

\bibitem{Kawamura:2023zuo}
J.~Kawamura and S.~Shin,
\href{doi:10.1007/JHEP11(2023)025}{JHEP \textbf{11}, 025 (2023)}.
[arXiv:2308.07814 [hep-ph]].

\bibitem{Bigaran:2023ris}
I.~Bigaran, B.~A.~Dobrescu and A.~Russo,
\href{doi:10.1103/PhysRevD.109.055033}{Phys. Rev. D \textbf{109}, no.5, 055033 (2024)}.
[arXiv:2312.09189 [hep-ph]].

\bibitem{ATLAS:2015qoy}
G.~Aad \textit{et al.} [ATLAS],
\href{doi:10.1007/JHEP09(2015)108}{JHEP \textbf{09}, 108 (2015)}.
[arXiv:1506.01291 [hep-ex]].

\bibitem{ATLAS:2023sbu}
G.~Aad \textit{et al.} [ATLAS],
\href{doi:10.1007/JHEP07(2023)118}{JHEP \textbf{07}, 118 (2023)}.
[arXiv:2303.05441 [hep-ex]].

\bibitem{CMS:2019hsm}
A.~M.~Sirunyan \textit{et al.} [CMS],
\href{doi:10.1103/PhysRevD.100.052003}{Phys.\ Rev.\ D \textbf{100}, 052003 (2019)}.
[arXiv:1905.10853 [hep-ex]].

\bibitem{CMS:2022nty}
A.~Tumasyan \textit{et al.} [CMS],
\href{doi:10.1103/PhysRevD.105.112007}{Phys.\ Rev.\ D \textbf{105}, 112007 (2022)}.
[arXiv:2202.08676 [hep-ex]].

\bibitem{CMS:2022cpe}
A.~Tumasyan \textit{et al.} [CMS],
\href{doi:10.1016/j.physletb.2023.137713}{Phys.\ Lett.\ B \textbf{846}, 137713 (2023)}.
[arXiv:2208.09700 [hep-ex]].

\bibitem{CMS:2024bni}
A.~Hayrapetyan \textit{et al.} [CMS],
\href{doi:10.1016/j.physrep.2024.09.012}{Phys.\ Rep.\ \textbf{1115}, 570-677 (2025)}.
[arXiv:2405.17605 [hep-ex]].

\bibitem{ATLAS:2024mrr}
G.~Aad \textit{et al.} [ATLAS],
\href{doi:10.1007/JHEP05(2025)075}{JHEP \textbf{05}, 075 (2025)}.
[arXiv:2411.07143 [hep-ex]].

\bibitem{Bissmann:2020lge}
S.~Bi{\ss}mann, G.~Hiller, C.~Hormigos-Feliu and D.~F.~Litim,
\href{doi:10.1140/epjc/s10052-021-08886-3}{Eur.\ Phys.\ J.\ C \textbf{81}, 101 (2021)}.
[arXiv:2011.12964 [hep-ph]].

\bibitem{Bhattiprolu:2019vdu}
P.~N.~Bhattiprolu and S.~P.~Martin,
\href{doi:10.1103/PhysRevD.100.015033}{Phys.\ Rev.\ D \textbf{100}, 015033 (2019)}.
[arXiv:1905.00498 [hep-ph]].

\bibitem{ILC:2013jhg}
H.~Baer \textit{et al.} [ILC],
[arXiv:1306.6352 [hep-ph]].

\bibitem{ILCInternationalDevelopmentTeam:2022izu}
A.~Aryshev \textit{et al.} [ILC International Development Team],
[arXiv:2203.07622 [physics.acc-ph]].

\bibitem{CLICDetector:2013tfe}
H.~Abramowicz \textit{et al.} [CLIC Detector and Physics Study],
[arXiv:1307.5288 [hep-ex]].

\bibitem{Franceschini:2019zsg}
R.~Franceschini,
\href{doi:10.1142/S0217751X20410158}{Int.\ J.\ Mod.\ Phys.\ A \textbf{35}, 2041015 (2020)}.
[arXiv:1902.10125 [hep-ph]].



\bibitem{Yang:2021dtc}
B.~Yang, J.~Li, M.~Wang and L.~Shang,
\href{doi:10.1103/PhysRevD.104.055019}{Phys.\ Rev.\ D \textbf{104}, 055019 (2021)}.

\bibitem{Shang:2021mgn}
L.~Shang, M.~Wang, Z.~Heng and B.~Yang,
\href{doi:10.1140/epjc/s10052-021-09152-2}{Eur.\ Phys.\ J.\ C \textbf{81}, 415 (2021)}.

\bibitem{Bhattacharya:2021ltd}
S.~Bhattacharya, S.~Jahedi and J.~Wudka,
\href{doi:10.1007/JHEP05(2022)009}{JHEP \textbf{05}, 009 (2022)}.
[arXiv:2106.02846 [hep-ph]].

\bibitem{Shang:2023rfv}
L.~Shang, J.~Li, X.~Jia and B.~Yang,
\href{doi:10.1016/j.nuclphysb.2022.116071}{Nucl.\ Phys.\ B \textbf{987}, 116071 (2023)}.

\bibitem{Bhattiprolu:2023yxa}
P.~N.~Bhattiprolu, S.~P.~Martin and A.~Pierce,
\href{doi:10.1103/PhysRevD.109.035009}{Phys.\ Rev.\ D \textbf{109}, 035009 (2024)}.
[arXiv:2308.08386 [hep-ph]].

\bibitem{Yue:2024sds}
C.~X.~Yue, Y.~Q.~Wang, H.~Wang, Y.~H.~Wang and S.~Li,
\href{doi:10.1016/j.nuclphysb.2024.116482}{Nucl.\ Phys.\ B \textbf{1000}, 116482 (2024)}.
[arXiv:2402.02072 [hep-ph]].




\bibitem{Shen:2025mxe}
J.~F.~Shen, Y.~J.~Zhang and L.~Han,
\href{doi:10.1088/1674-1137/adf1ef}{Chin. Phys. C \textbf{49}, no.11, 113108 (2025)}.
\bibitem{Liu:2025akp}
Y.~B.~Liu and J.~W.~Lian,
\href{doi:10.1088/1674-1137/ae2d24}{Chin. Phys. C \textbf{50}, no.3, 033110 (2026)}.
[arXiv:2511.02557 [hep-ph]].


\bibitem{Sagheer:2026eql}
H.~Sagheer, I.~Ahmed and J.~Muhammad,
[arXiv:2606.28717 [hep-ph]].
\bibitem{Liu:2025ori}
Y.~B.~Liu and S.~Moretti,
\href{doi:10.1103/n8gp-dcx6}{Phys. Rev. D \textbf{113}, no.5, 055034 (2026)}.
[arXiv:2512.22490 [hep-ph]].
\bibitem{Das:2020gnt}
A.~Das, S.~Mandal and T.~Modak,
\href{doi:10.1103/PhysRevD.102.033001}{Phys. Rev. D \textbf{102}, no.3, 033001 (2020)}.
[arXiv:2005.02267 [hep-ph]].

\bibitem{Das:2020uer}
A.~Das and S.~Mandal,
\href{doi:10.1016/j.nuclphysb.2021.115374}{Nucl. Phys. B \textbf{966}, 115374 (2021)}.
[arXiv:2006.04123 [hep-ph]].

\bibitem{Yue:2024ftz}
C.~X.~Yue, Y.~Q.~Wang, X.~C.~Sun and X.~Y.~Li,
\href{doi:10.1088/1361-6471/ad9ec8}{J. Phys. G \textbf{52}, no.2, 025003 (2025)}.
[arXiv:2412.07125 [hep-ph]].

\bibitem{Yue:2025xvk}
C.~X.~Yue, M.~S.~Y.~Wang and X.~Y.~Li,
\href{doi:10.1016/j.nuclphysb.2025.117143}{Nucl. Phys. B \textbf{1020}, 117143 (2025)}.
\bibitem{Li:2025epjc}
R.~P.~Li, J.~W.~Lian and Y.~B. Liu,
\href{doi:10.1140/epjc/s10052-025-15178-7}{Eur.\ Phys.\ J.\ C \textbf{85}, 1429 (2025)}.


\bibitem{deBlas:2013gla}
J.~de Blas,
\href{doi:10.1051/epjconf/20136019008}{EPJ Web Conf. \textbf{60}, 19008 (2013)}.
[arXiv:1307.6173 [hep-ph]].
\bibitem{Cui:2026wwo}
Y.~Cui, S.~Wang, Z.~H.~Yu and H.~H.~Zhang,
\href{doi:10.1103/p8l7-vv3p}{Phys. Rev. D \textbf{113}, no.11, 115025 (2026)}.
[arXiv:2604.11232 [hep-ph]].
\bibitem{Cynolter:2008ea}
G.~Cynolter and E.~Lendvai,
\href{doi:10.1140/epjc/s10052-008-0771-7}{Eur. Phys. J. C \textbf{58}, 463-469 (2008)}.
[arXiv:0804.4080 [hep-ph]].




\bibitem{Adhikary:2024esf}
A.~Adhikary, M.~Olechowski, J.~Rosiek and M.~Ryczkowski,
\href{doi:10.1103/PhysRevD.110.075029}{Phys. Rev. D \textbf{110}, no.7, 075029  (2024)}.
[arXiv:2406.16050 [hep-ph]].

\bibitem{deBlas:2025pco}
J.~de Blas, C.~Giuliano, G.~Guedes, R.~S.~L{\'o}pez and J.~Santiago,
\href{doi:10.1007/JHEP05(2026)142}{JHEP \textbf{05}, 142 (2026)}.
[arXiv:2512.04148 [hep-ph]].

\bibitem{Daberstiel:2026wxe}
G.~Daberstiel, K.~M{\"o}hling, D.~St{\"o}ckinger and H.~St{\"o}ckinger-Kim,
\href{doi:10.1007/JHEP07(2026)259}{JHEP \textbf{07}, 259 (2026)}.
[arXiv:2603.21414 [hep-ph]].




\bibitem{ATLAS:2019gqq}
M.~Aaboud \textit{et al.} [ATLAS],
\href{doi:10.1103/PhysRevD.99.092007}{Phys.\ Rev.\ D \textbf{99}, 092007 (2019)}.
[arXiv:1902.01636 [hep-ex]].

\bibitem{Alwall:2014hca}
J.~Alwall, R.~Frederix, S.~Frixione, V.~Hirschi, F.~Maltoni, O.~Mattelaer, H.-S.~Shao, T.~Stelzer, P.~Torrielli and M.~Zaro,
\href{doi:10.1007/JHEP07(2014)079}{JHEP \textbf{07}, 079 (2014)}.
[arXiv:1405.0301 [hep-ph]].

\bibitem{Sjostrand:2014zea}
T.~Sj\"ostrand, S.~Ask, J.~R.~Christiansen \textit{et al.},
\href{doi:10.1016/j.cpc.2015.01.024}{Comput.\ Phys.\ Commun.\ \textbf{191}, 159 (2015)}.
[arXiv:1410.3012 [hep-ph]].

\bibitem{deFavereau:2013fsa}
J.~de Favereau \textit{et al.} [DELPHES 3],
\href{doi:10.1007/JHEP02(2014)057}{JHEP \textbf{02}, 057 (2014)}.
[arXiv:1307.6346 [hep-ex]].

\bibitem{Behnke:2013lya}
T.~Behnke, J.~E.~Brau, P.~N.~Burrows, J.~Fuster, M.~Peskin, M.~Stanitzki, Y.~Sugimoto, S.~Yamada, H.~Yamamoto and H.~Abramowicz, \textit{et al.}
[arXiv:1306.6329 [physics.ins-det]].

\bibitem{ILDConceptGroup:2020sfq}
H.~Abramowicz \textit{et al.} [ILD Concept Group],
[arXiv:2003.01116 [physics.ins-det]].

\bibitem{CMS:2012feb}
S.~Chatrchyan \textit{et al.} [CMS],
\href{doi:10.1088/1748-0221/8/04/P04013}{JINST \textbf{8}, P04013 (2013)}.
[arXiv:1211.4462 [hep-ex]].

\bibitem{Dokshitzer:1997in}
Y.~L.~Dokshitzer, G.~D.~Leder, S.~Moretti and B.~R.~Webber,
\href{doi:10.1088/1126-6708/1997/08/001}{JHEP \textbf{08}, 001 (1997)}.



\bibitem{Wobisch:1998wt}
M.~Wobisch and T.~Wengler,
arXiv:hep-ph/9907280.

\bibitem{Cacciari:2011ma}
M.~Cacciari, G.~P.~Salam and G.~Soyez,
\href{doi:10.1140/epjc/s10052-012-1896-2}{Eur. Phys. J. C \textbf{72}, 1896 (2012)}.



\bibitem{ma5}
E.~Conte, B.~Fuks and G.~Serret,
\href{doi:10.1016/j.cpc.2012.09.009}{Comput.\ Phys.\ Commun.\ \textbf{184}, 222 (2013)}.
[arXiv:1206.1599 [hep-ph]].

\bibitem{Conte:2014zja}
E.~Conte, B.~Dumont, B.~Fuks and C.~Wymant,
\href{doi:10.1140/epjc/s10052-014-3103-0}{Eur.\ Phys.\ J.\ C \textbf{74}, 3103 (2014)}.
[arXiv:1405.3982 [hep-ph]].

\bibitem{Cowan:2010js}
G.~Cowan, K.~Cranmer, E.~Gross and O.~Vitells,
\href{doi:10.1140/epjc/s10052-011-1554-0}{Eur.\ Phys.\ J.\ C \textbf{71}, 1554 (2011)}.
[Erratum: Eur.\ Phys.\ J.\ C \textbf{73}, 2501 (2013)].
[arXiv:1007.1727 [physics.data-an]].

\end{thebibliography}
\end{document}